\documentclass[useAMS,usenatbib,usegraphicx,dvips]{mn2e}
\usepackage{times}
\usepackage{longtable}
\usepackage{url}
\title{Multicolour CCD photometry of the variable stars in globular 
cluster M3}
\author[J.~M. Benk\H{o} et al.]
{J.~M.~Benk\H{o}$^{1}$\thanks{E-mail: benko@konkoly.hu},
G. \'A.~Bakos,$^{1,2,3}$ and J. Nuspl$^{1}$\\
$^{1}$Konkoly Observatory of the Hungarian Academy of Sciences,
        P O Box 67, H-1525 Budapest, Hungary \\
$^{2}$Harvard-Smithsonian Center for Astrophysics, 60 Garden Street,
Cambridge, MA02138, USA \\
$^{3}$Hubble Fellow
}
\date{Accepted 2006 August 18. Received 2006 August 11; 
      in original form 2006 June 27}

\pagerange{\pageref{firstpage}--\pageref{lastpage}}
\pubyear{2006}

\begin{document}

\maketitle

\label{firstpage}

\begin{abstract}
We present time series data on the variable stars of the 
galactic globular cluster Messier~3 (M3).
We give {\it BVI$_{\rm C}$\/} light curves for 226 RR Lyrae, 2 SX
Phe and 1 W Vir type variables, along with estimated fundamental photometric
parameters such as intensity and magnitude-averaged brightness and
pulsation periods. In some cases the periods we have found
significantly differ from the previously published ones.
This is the first published light curve and period determination for
variable V266.
The {\it I\/}-band light curve has not been observed previously 
for numerous (76) variables.
Three new RR Lyrae variables have been discovered.
Groups of RR Lyrae variables that belong to different evolutionary stages
and have been separated previously on the basis of {\it V\/} data were
found here for all colours and colour indices by cluster analysis.
The {\it I\/}-band period \--- luminosity relation is also discussed. From
the 66 modulated (Blazhko type) RR Lyrae stars we investigated, six are
newly identified and two of them are first overtone pulsators. In the
case of 13 RR Lyrae, the period of Blazhko cycle has been estimated for
the first time.  V252 is identified as a new RRd variable.
Amplitude ratio of RRd stars have been investigated to search
possible mode content changes. In contrast to previous
publications no changes have been found.
Problems with the sampling of the time series of
typical cluster variability surveys is demonstrated.
\end{abstract}


\section{Introduction}

Multicolour photometry of RR Lyrae stars is very important for obtaining
information on their physical parameters such as temperature, mass and
luminosity. RR Lyrae stars in globular clusters deserve additional
interest because they offer an excellent test for the stellar evolution
theory and cosmological distance scale.  Regular monitoring of clusters'
variables is also essential to investigate various subtle effects, such
as period
and modal content changes of multimode RR Lyrae stars, or the Blazhko
phenomenon.

M3 (NGC~5272;
$\alpha_{2000}=13^{\rm h}42^{\rm m}11\fs 2$,
$\delta_{2000}=28\degr 22\arcmin 32\arcsec $)
has been target of a large number of different studies. Although it is
the most variable-rich galactic globular cluster (with 274 catalogued
variables,
with most of these belonging to the RR Lyrae type), the number of high
quality time series
observations are still limited. \citet{CC01} (hereafter CC01) published
a comparable amount of data in {\it B\/} and {\it V\/} colours to the
dataset presented here.
In addition, only three CCD-observations based publications exist with
partial coverage of M3: \citet{K98} (K98) in {\it V}-band, \citet{C98}
(C98) and \citet{hartman} (H05) in {\it BVI}. \citet*{Strader} (S02)
derived {\it V\/} time series for stars in the core of the cluster, but
their light curves represent differential fluxes on an arbitrary scale,
instead of standard magnitudes.

The lack of data on M3 inspired several time series studies; such as
\citet*{Mallik}, \citet{Ug} and \citet{AFer}.
The results of these investigations have not been published yet.

We started a multicolour {\it BVI$_{\rm C}$\/} CCD survey of globular
clusters at Konkoly Observatory in 1998.  The main goal of the project
is to obtain high quality time series for as many globular cluster
variables as possible. The instrumentation enables us to observe the
clusters simultaneously with two telescopes (1-m~RCC and 90/60-cm
Schmidt), both equipped with CCD cameras. Their different fields of
view (FOVs) make it possible to obtain images with sufficient
resolution for accurate photometry of both the crowded central regions
and the dispersed halos, respectively.  In this work we present the
observations of M3.

\begin{figure}
\includegraphics[width=8.5cm]{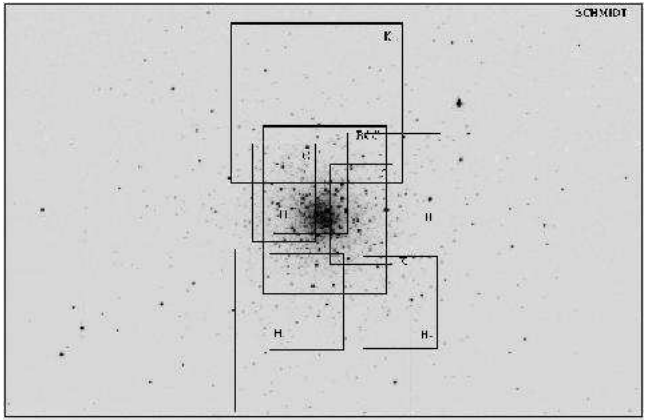}
\caption[]{The relative sizes and positions of the frames used in the
 published CCD time series papers on M3. The notations are:
`K' \--- \citet{K98}, `C' \--- \citet{C98}, \--- `H'
\citet{hartman}, \---`RCC' + `SCHMIDT' \--- present work. \citet{CC01}
have used a slightly larger FOV than `SCHMIDT'.}
\label{frames}
\end{figure}

\section{Observations and data reduction}

\subsection{Observations}

The observational material of M3 was obtained in two seasons (1998 and
1999) with two telescopes (see Table~\ref{obs} for details).  Both
telescopes are mounted at Piszk\'estet\H{o} Mountain Station of the
Konkoly Observatory.
The 60/90/180-cm Schmidt telescope is equipped with a CCD camera by
Photometrics Inc.~having a Kodak KAF-1600
$1024\times1536$ chip. The setup yields a $19'\times28'$ FOV, with $1\farcs
0$/pixel resolution.
The 1-m RCC telescope was used with two cameras.
In 1998 a camera constructed by Wright Instruments was attached to
the Cassegrain focus. This device contains an EEV
CCD05-20 $800\times1200$ chip that corresponds to a $4'\times6'$ FOV with
$0\farcs35$/pixel resolution. In 1999 a Photometrics camera was used.
Its basic parameters are: Thomson
7896M 1k$\times$1k chip, $5'\times5'$ FOV with $0\farcs29$ resolution.
For the calibration process of the cameras and other technical details
see \citet{calib}. Standard Johnson {\it BV\/} and Kron-Cousins
{\it I$_{\rm C}$\/} filters were used for all observations. From now on the
band-index $C$ is generally omitted.

The total number of images are 521 in {\it V}, 226 in {\it B\/} and
384 in {\it I}.  Sky flats (or alternatively dome flats) were taken
as main calibration images.  Each night some bias frames (5-50) and
dark images (with 3-5 minutes exposure times) were also taken.

The relative positions for the fields-of-view and the cluster for previous
studies and this work are shown in Fig.~\ref{frames}.  The FOV of our
frames allowed us to get light curves from almost all variables of the
cluster. The nomenclature of variables used here are the same as in the
General Catalogue of Variable Stars in Globular Clusters (\citealt{Cl},
\--- hereafter The Catalogue).

\begin{table}
\caption{Log of the observations of M3.}
\begin{tabular}{c@{}p{6mm}*5{p{6mm}}l}
\hline
Date  & \multicolumn{6}{c}{Filter / Exp.~time [sec]} &  No. \\

[JD-2400000] &  \multicolumn{3}{c}{Schmidt} & \multicolumn{3}{c}{RCC} &\\
\hline
50893 &    & $V/300$ &    &  & & & 22\\
50894 &    & $V/300$ &    &  & & & 35\\
50896 &    & $V/300$ &    & $B/300$ & $V/300$ & $I/180$ & 54\\
50897 &   & $V/300$ &    &  & $V/240$ &  $I/180$  & 67\\
50898 &    &         &    &  & $V/240$ &  $I/180$    & 27 \\
50941 &    &         &    & $B/300$ & $V/240$ & $I/180$ & 51    \\
50942 &    &         &    & $B/300$ & $V/240$ & $I/180$  & 56  \\
50943 & $B/300$ & $V/240$ & $I/180$ & $B/300$ & $V/240$ & $I/180$ &135\\
50944 & $B/300$ & $V/180$ & $I/80$  & $B/300$ & $V/240$ & $I/180$  &117\\
50960 & $B/300$ & $V/240$ & $I/180$ &   &  & & 36\\
50961 & $B/300$ & $V/240$ & $I/180$ &   &  & & 60\\
50970 & $B/300$ & $V/240$ & $I/180$ &   &  & & 46\\
50971 & $B/100$ & $V/80$  & $I/60$  &   &  & & 51$^*$\\
50972 & $B/100$ & $V/80$  & $I/60$  &   &  & & 31$^*$\\
51256 &   & &                       & $B/240$ & $V/180$ & $I/180$  &75 \\
51257 &   & &                       &    &      $V/180$ & $I/180$  &50 \\
51259 &   & &                       &    &      $V/180$ & $I/180$  &37 \\
51262 &   & &                       &    &      $V/240$ & $I/240$  &78 \\
51283 &   & &                       & $B/240$ & $V/180$ & $I/180$  &69 \\
\hline
\end{tabular}

\footnotesize{$^*$ Three images were combined
for each data point.}

\label{obs}
\end{table}

\subsection{Data reduction \& photometry}\label{red}

We used the {\sc iraf/ccdred}{\footnote{{\sc iraf} is distributed by the NOAO,
operated by the Association of Universities for Research in Astronomy
Inc., under contract with the NSF.}} package for the standard
reduction procedures: bias, dark and flat field correction. Other
corrections (e.g.~deferred charge, non-linearity) were also
investigated, but they were found to be unjustified to apply.

The brightness of stars were determined first by using the aperture
photometry task {\sc daophot/phot} of {\sc Iraf}.  This method
provides robust flux estimates with small errors for isolated stars,
and were used for variables that lie
in the outer parts of Schmidt frames ({\it method A}\/).

\begin{figure*}
\includegraphics[width=15cm]{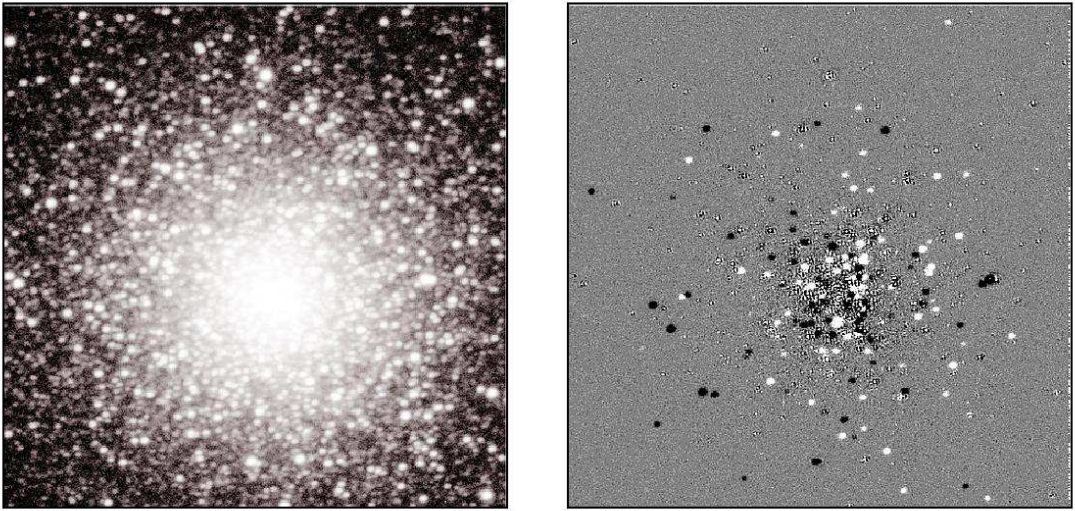}
\caption[]{A typical result of the subtraction process applying the ISM
method. The left panel shows the most crowded central part of a {\it
V\/} frame of the Schmidt telescope, and the right panel shows the same
area after subtraction.  It is easy to identify the variables (white
and black spots depending on whether they are brighter or fainter
compared to the reference image). The other stars practically
disappeared, except for few saturated ones.}
\label{sub}
\end{figure*}

\subsubsection{Image subtraction photometry}

In the case of  variable star photometry on globular clusters
the Image Subtraction Method (ISM) (\citealt{AlaLup}, \citealt{Al2}) is
more efficient and accurate than the traditional PSF fitting methods,
such as {\sc dophot} \citep*{DoP} or {\sc daophot} \citep{daophot}:
more variable stars can be measured in the dense
regions, and the r.m.s.~of the light curves is also significantly
better.
We used different versions of the Alard's {\sc Isis} software
package{\footnote{Available via
\url{http://www.iap.fr/users/alard/package.html}.}} 
as an implementation of ISM. In practice, the program
was run several times by varying the most important parameters ({\tt
nstamps\_x, nstamps\_y, half\_mesh\_size, half\_stamp\_size,
saturation, deg\_spatial}) to obtain optimal subtractions.
Fig.~\ref{sub} demonstrates the effectiveness of the method by showing
an original and a subtracted frame.

Different versions of {\sc Isis} occasionally resulted in bad
photometry for several stars. Typical errors were: apparent hump(s)
in the light curve, smaller amplitude than expected
(amplitude errors), or unexpectedly high r.m.s.  To
check on these problems the {\sc Iraf/phot} task has been run on the
subtracted images. Fortunately, this simple trick has solved almost all
trouble. Hence this type of photometry was also used alternatively.

The ISM method gives differential fluxes of objects with respect to their
fluxes measured on the reference frame.  For the instrumental magnitudes we need
to determine the magnitudes and fluxes for each variable in the
reference frames as accurately as possible.  To get these zero points,
we performed PSF photometry using {\sc iraf/daophot} with variable PSF for
the photometric reference images (one for each colour and camera).  The
problems of the transformation to magnitude scale have been discussed
elsewhere \citep{Ben}. The zero-points were derived
from either the RCC telescope's data ({\it method R}, in case the star was
within the RCC field boundaries) or from the Schmidt frames, when it was not
measured by RCC telescope ({\it method S}\/).

In the case of variables that are both on the RCC and Schmidt frames, RCC
telescope data were used as a reference and the Schmidt telescope data
were transformed to the RCC.  In order to compute the proper zero point
values the two light curves were cross-correlated by a 2D
Kolmogorov\---Smirnov algorithm (see \citealt{NumRec}).  For stars with
variable light curve (e.g.~Blazhko effect, RRd type) light curve parts of
simultaneous nights (1998 May 9 \& 10) were used.

The three type of photometries (A, R, S) were tested whether
they give homogeneous results or not. Our average magnitudes (see
Sec.~\ref{Avbr}) were cross-correlated with CC01's ones for the
stars in common. For a given colour all three methods yielded the
same linear correlation: parameters of the least-squares fits
agree within the regression error.

\subsubsection{Transformation into the standard system}

The instrumental magnitudes were transformed into the  
{\it BVI$_{\rm C}$\/} Johnson-Cousins system.  
The standard sequence published by
\citet{St3} was used as a source of standard magnitudes.
Stars that are not well separated in our frames have been
omitted from these data sets. Even so, dozens of standards remained for
all bands.  Let us consider the colour equations as

\begin{eqnarray}\label{ce}
V&=&m_v-k'_VX+\varepsilon(B-V)+\xi_V, \nonumber \\
B&=&m_b-k'_BX+\mu(B-V)+\xi_B, \\
I&=&m_i-k'_IX+i(V-I)+\xi_I, \nonumber
\end{eqnarray}
where the designations are the usual: {\it B}, {\it V}, {\it I\/}
are the transformed and $m_v$, $m_b$, $m_i$ are the instrumental magnitudes,
$k'$s are the extinction coefficients, and $X$ is the air mass,
$\varepsilon$, $\mu$, $i$ are the telescope constants and $\xi$s are
the zero points.  The telescope constants for the reference images have
been determined by linear least-square fitting.

\begin{table}
\caption{The determined telescope constants. Coefficients are defined by Eq.
\ref{ce}. The errors are the 1-$\sigma$ of the linear
least-square fits.}
\begin{tabular}{@{}lccc}
\hline
 System      & $\varepsilon$             & $\mu$           & $i$  \\
\hline
 RCC '98     & $\phantom{-}0.003\pm0.01$ & $\phantom{-}0.202\pm0.01$ &
$-0.040\pm0.02$ \\
Schmidt '98 & $\phantom{-}0.008\pm0.07$ & $-0.080\pm0.08$ &
$-0.016\pm0.06$ \\
RCC '99     & $\phantom{-}0.004\pm0.02$ & $\phantom{-}0.046\pm0.02$ &
$-0.038\pm0.03$ \\
\hline
\end{tabular}
\label{const}
\end{table}

As the telescope constant $\varepsilon$ in all cases
turned out to be zero within
the error all coefficients were calculated with the value of
$\varepsilon=0$.  The obtained values are shown in the Table~\ref{const}.
Typical error of an individual
measurement is between $0.01$ \--- $0.02$ mag in
the bands {\it V\/} and {\it I\/} and $0.03$ \--- $0.04$ mag in
{\it B}.

\section{Analysis and Results}

\subsection{Parameters of the light curves}\label{parlight}

We are presenting data here on 238 variable stars, three of them are new
discoveries.  The light curves of V266 are given for the first time.
 The construction of our observational database{\footnote{
Available via \url{http://www.konkoly.hu/staff/benko/pub.html}.}} 
is simple:
each star has three separate files including the heliocentric Julian
date (HJD) and magnitudes in the Johnson-Cousins reference
{\it B}, {\it V\/} and {\it I}, respectively.

\begin{table}
\caption{Non-variable stars with variable ID in the Catalogue.}
\centering
\begin{tabular}{@{}lccc}
\hline
ID & B & V & I \\
\hline
2	&	15.874	&	15.189	&	14.143\\
98	&	15.387	&	15.593	&	13.327\\
102	&	16.121	&	15.674	&	14.986\\ 
103	&	16.565	&	15.930	&	15.026\\
153	&	14.562	&	13.574	&	12.307\\
164	&	14.730	&	13.818	&	12.633\\
169	&	15.273	&	14.864	&	13.641\\
182	&	15.967	&	15.454	&	14.726\\
204	&	15.948	&	15.877	&	15.042\\
227	&	15.995	&	15.469	&	14.769\\
228	&	16.016	&	15.630	&	15.084\\
231	&	16.074	&	16.000	&	15.851\\
232	&	15.950	&	15.852	&	15.885\\
233	&	16.224	&	15.935	&	14.680\\

\hline
\end{tabular}
\label{nonvar}
\end{table}

\begin{figure}
\includegraphics[width=8.5cm]{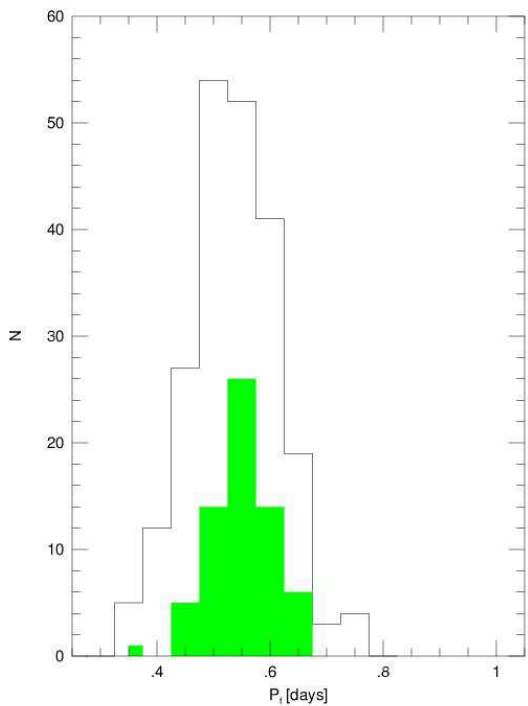}
\caption[]{
Histogram of fundamentalized periods ($P_{\rm f}$) of M3 RR Lyrae variable
stars. Shaded bars show modulated (Blazhko type) variables which are 
mostly responsible for peaked distribution. 
}
\label{period_distrib}
\end{figure}

\subsubsection{Periods}\label{peri}

Each variable star was analyzed to find its period by using a period search
on the {\it V\/} data. Discrete Fourier Transformation was
applied by using the facilities of the program package {\sc mufran}
\citep{Zoli} or a Phase Dispersion Minimization ({\sc iraf/pdm},
\citealt{Stel}) depending on the time distribution of collected data.
Periods of the literature were accepted where the newly determined
periods do not differ significantly from the old ones.  The {\it V\/}
phase diagrams are shown in Fig.~\ref{light_curves}.

In the period search, outlier data points were omitted from the light
curves manually.  Generally, those isolated points were rejected that
deviated by more than 3$\sigma$ from the fitted curve.  In some cases,
when a variable is lying near to a bright star, complete nights have to
be excluded if the neighbouring bright star was saturated and
therefore the ISM did not work well.

In the figure \ref{period_distrib} we show ``fundamentalized'' 
period (for definition see \citealt{vAB}) 
distribution from our (re)calculated periods.
The peaked period distribution of M3 was already obvious in figure of
\citet{Oo} but only \citet{CT} and \citet{RC} revealed that it
is an evidence of peculiar distribution of stars within 
the instability strip. \citet{Cat04} has reached the conclusion that
this peaked period distribution might pose troubles for conventional
stellar evolution theory. Assuming a bimodal mass distribution 
\citet*{Castellani} obtained reasonable agreement 
between standard evolutionary model and observed period distribution.
Therefore they do not believe in serious problems in connection with 
canonical models and rather prefer Catelan's suggestion:
``M3 might be a pathological case that cannot be considered 
representative of the Oosterhoff~I class''.
Although the evolutionary theory is beyond the scope of this paper
we add further argument to latter statement in Sec.~\ref{cmdsec}. 

\subsubsection{Average brightnesses}\label{Avbr}

Following this, a least-square fit of a Fourier sum containing
5 \--- 15 sine terms was calculated.

Mean magnitudes were derived by averaging over the
pulsation cycle of these fitted light curves both in intensity (then
converted back to magnitude) and magnitude scales.
Throughout this paper, the  magnitude-averaged mean 
brightness is denoted by
$\overline{B}$, $\overline{V}$ and
$\overline{I}$, and
$\langle B \rangle$, $\langle V \rangle$ and
$\langle I \rangle$ symbols indicate intensity means.

This process, however, might result in erroneous values for RR Lyrae stars
with Blazhko effect because Blazhko cycles have no complete coverage.
Investigating Blazhko type RR Lyrae in LMC \citet{Alcock} have
concluded that the average brightnesses varied by less than 0.006 mag
over the whole Blazhko cycle.  Therefore, using some subsequent nights'
observations, well covered phase diagrams had been constructed for each
Blazhko type RR Lyrae star, and then the average brightnesses were
calculated from these phase diagrams.  For about a dozen stars there
was more than one possibility to make a good phase diagram. In these
cases the average magnitudes were calculated from each diagram
separately. By comparing the mean magnitudes of the same star at different
Blazhko phases we found that the values are the same within 0.01 mag.

\subsubsection{Classification}

We measured 226 variable stars classified as RR Lyraes,
169 out of these turned out to be fundamental mode pulsators
(RRab) and 48 to be first overtone pulsators (RRc). Nine variables have been
found to pulsate in two modes simultaneously (RRd). We have found 64
amplitude and/or phase modulated (Blazhko effect) stars among RRab-s
and 2 among RRc-s. The incidence rates of Blazhko variables are 36.7
and 4.2 per cent for RRab and RRc stars, respectively. The rate of
modulated RRc variable stars is in good agreement with the values 4 \---
6 per cent given by all different studies on large surveys
(\citealt{Alcock2000}, \citealt{Moskalik},
\citealt{Soszy}). Our rate, calculated from RRab stars agrees well with
that of \citet{Moskalik} as determined from the OGLE~I data of the
Galactic bulge.
Measurements in LMC resulted in a much lower
rates: 11.9 per cent from MACHO data \citep{Alcock} and
15 per cent from OGLE~II survey \citep{Soszy}, respectively.

\begin{figure}
\includegraphics[width=8.5cm]{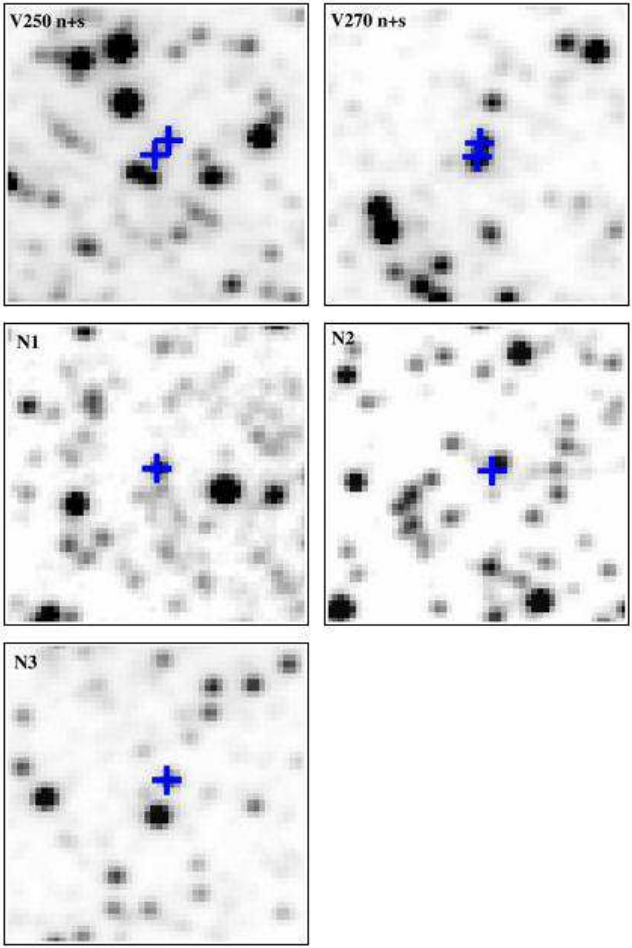}
\caption[]{Finding charts for non-catalogued variables. Boxes are
$\approx15''$ in width. North is up, East is to the left.}
\label{find}
\end{figure}

\begin{table*}
\begin{minipage}{170mm}
\caption{The basic parameters of the variables in M3.
Table~\ref{table4} is published in its entirety in the electronic 
edition of this journal} \label{table4} 
\begin{tabular}{@{}*{11}{l}}
\hline 
ID & Period & Type &
$\overline{B} $ & $\langle B \rangle$ &
$\overline{V} $ & $\langle V \rangle$ & 
$\overline{I} $ & $\langle I \rangle$ &
 ${\rm Ref}_{B, V, I}$ &  Comm. \\
\hline
1	&	0.5205963	&	RRab	&	16.004	&	15.900	&	15.688	&	15.631	&	15.248	&	15.227	&	R	R	R	&	3\\ 
3	&	0.5581979	&	RRab	&	15.939	&	15.857	&	15.661	&	15.612	&	15.175	&	15.155	&	R	R	R	&	Bl,3\\ 
4n	&	0.585029	&	RRab	&	 	&	 	&	 	&	 	&	 	&	 	&	R	R	R	&	c\\ 
4s	&	0.593069	&	RRab	&	 	&	 	&	 	&	 	&	 	&	 	&	R	R	R	&	c\\ 
5	&	0.505703	&	RRab	&	15.986	&	15.974	&	15.725	&	15.714	&	15.243:	&	15.238:	&	A	A	S	&	Bl,c\\ 
6	&	0.5143327	&	RRab	&	16.083	&	16.003	&	15.755	&	15.703	&	15.294	&	15.276	&	R	R	R	&	1\\ 
7	&	0.4974248	&	RRab	&	15.980	&	15.889	&	15.694	&	15.638	&	15.248	&	15.230	&	R	R	R	&	Bl,2\\ 
8	&	0.636728	&	RRab	&	15.808	&	15.787	&	15.649	&	15.634	&	14.902	&	14.898	&	R:	R:	R:	&	3\\ 
9	&	0.5415553	&	RRab	&	16.031	&	15.957	&	15.689	&	15.645	&	15.233	&	15.217	&	A	A	S	&	2\\ 
10	&	0.5695465	&	RRab	&	16.057	&	16.004	&	15.684	&	15.647	&	15.162	&	15.152	&	S	S	S	&	Bl,2,c\\ 
11	&	0.5078915	&	RRab	&	15.993	&	15.873	&	15.706	&	15.636	&	15.188	&	15.166	&	S	S	S	&	3\\ 
12	&	0.3179347	&	RRc	&	15.840	&	15.814	&	15.621	&	15.606	&	15.281	&	15.275	&	R	R	R	&	 \\ 
13	&	0.4795043	&	RRd	&	15.931	&	15.917	&	15.691	&	15.684	&	15.243	&	15.239	&	R	R	R	&	$P_1=0.351579$\\ 
14	&	0.6359002	&	RRab	&	15.914	&	15.867	&	15.581	&	15.551	&	15.059	&	15.050	&	R	R	R	&	Bl,4\\ 
15	&	0.5300874	&	RRab	&	15.919	&	15.833	&	15.667	&	15.616	&	15.159	&	15.144	&	A	A	A	&	3\\ 

\hline
\end{tabular}
\end{minipage}
\end{table*}

\begin{figure*}
\includegraphics[width=18.5cm]{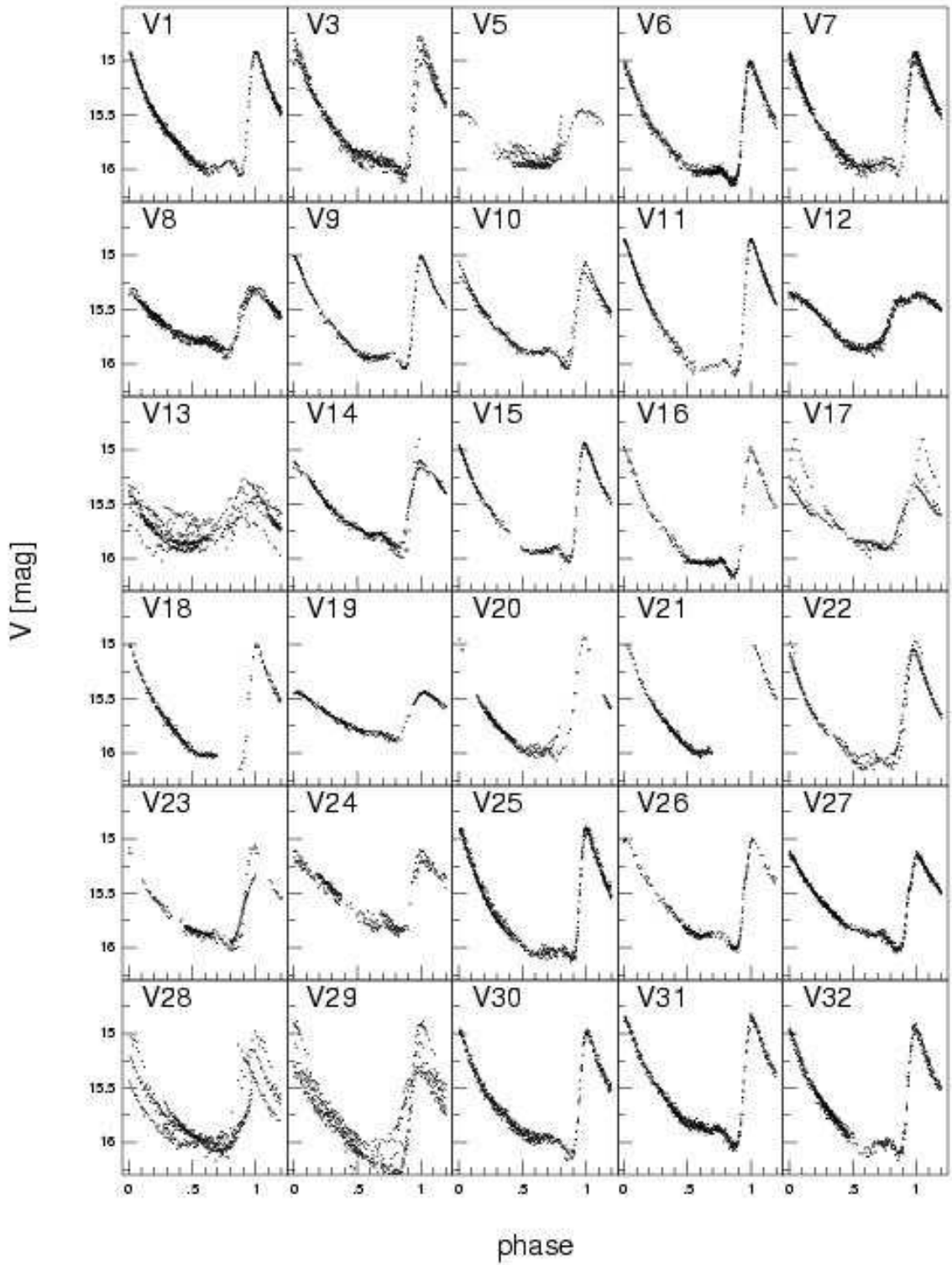}
\vspace*{-1.5cm}
\caption[]{The $V$ phase diagrams of the measured variables.
This figure is published in its entirety in the electronic 
edition of this journal}
\label{light_curves}
\end{figure*}

\begin{figure*}
\addtocounter{figure}{-1}
\includegraphics[width=18.5cm]{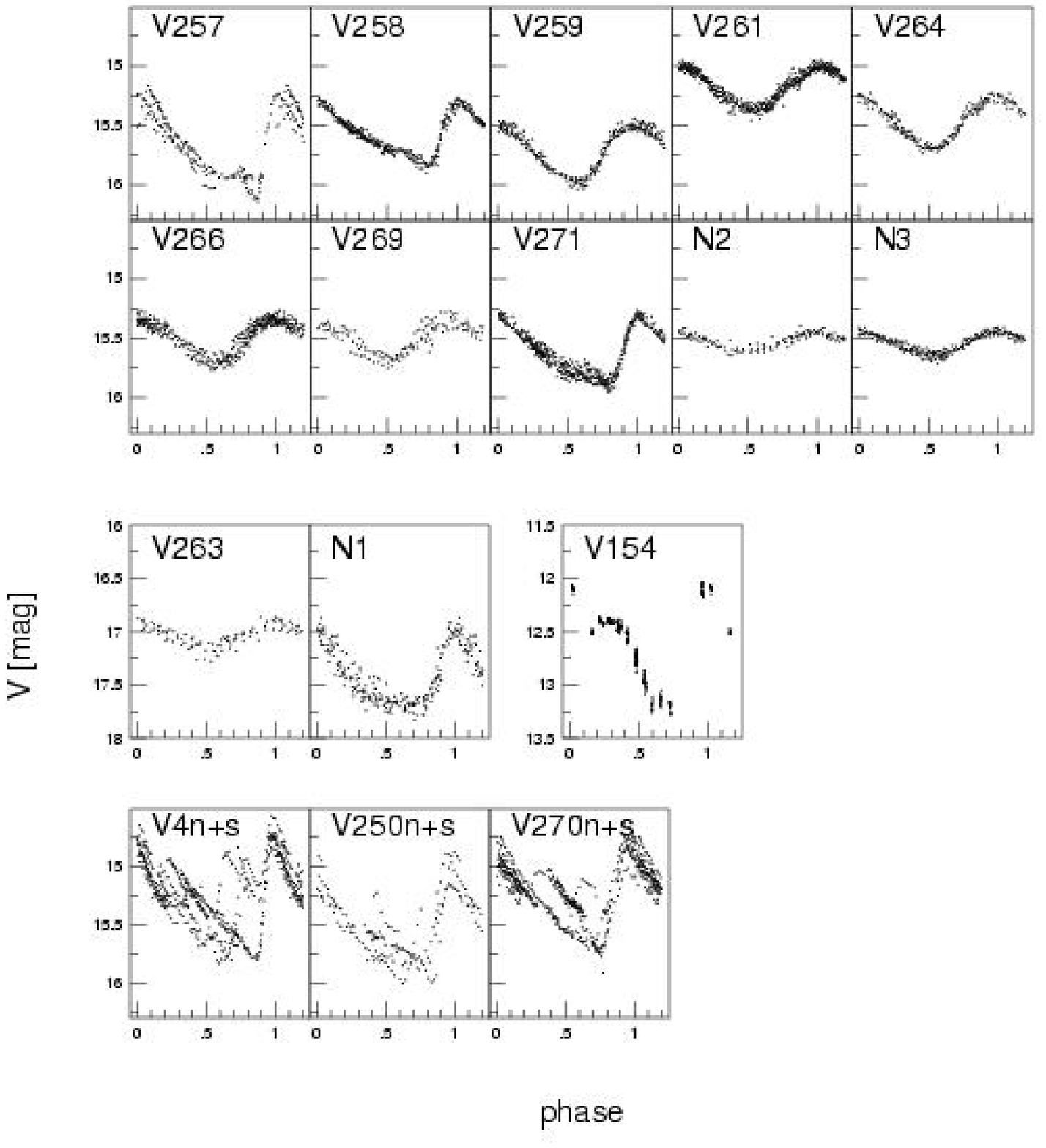}
\vspace*{-1.5cm}
\caption[]{(continued)}
\end{figure*}

\subsubsection{Catalogue of variables}

Basic parameters of the light curves are summarized in Table~\ref{table4}.
In column 1 the ID comes from The Catalogue.  Column 2
shows the pulsation period, column 3 the type of variability, columns
4\---9 the magnitude and intensity averages. Column 10 reference letters
indicate the type of magnitude transformation: `A', `S' and `R' (see
Sec.~\ref{red} for the details).  In the last column comment `Bl'
indicates Blazhko type variability, `m' merging with nearby variable or
a saturated bright star and `c' extra comments in the text.  The second
(non-dominant) period of double mode RR Lyrae is also shown in this
column. When our data have no complete phase coverage, the average
brightnesses were derived by using all published data shifted to our
ones. Comment `a' refers to this. The colon
at an average magnitude indicates noisy or gappy light curve and at a
reference letter it means a doubtful zero point of the magnitude
transformation.

Numerous stars can be found in The Catalogue, which turned out to be
non-variable. We summarized our observations on these constant stars in
Table~\ref{nonvar}.
It has to be mentioned that even the newest on-line version of
The Catalogue{\footnote{Available at 
\url{http://www.astro.utoronto.ca/~cclement/cat/listngc.html}.}} contains
position errors. The column declination must have been shifted
accidentally by one line at the variables V238, so the declinations of
V238, V239 are wrong. In the Table~\ref{pos} the correct positions are
repeated to avoid any further confusion.

\subsection{New variable stars}\label{new}

During our preliminary data processing six new variable stars have been
discovered and published elsewhere \citep{Acta}. We searched for
additional new variable stars in `variability images' prepared by {\sc
Isis}.  Scanning these images three additional
new variables have been identified.
They are designed as N1, 
N2 and N3 (for finding charts see Fig.~\ref{find}).
To verify their positions and light curve parameters
turned out that N1 and N3 are the same stars as NV291 and NV292,
respectively in the meantime published study of \citet{hartman}.  The
colour indices, period and light curve shapes suggest that N2 is a
new RRc type star.

When the images of two (or more) variable stars are blended their
light curves became confusing.
This was the situation in the case of the light curves of V250 and
V270 (see Fig.~\ref{light_curves}),
albeit no close companions were known to them. By
inspecting the subtracted images we have found several ones
in both cases in which two close variable stars can be separated.
The blended light curves suggest that in both cases two
RRab stars are merged.
We have split their notations into two as V250n,
V250s and V270n, V270s as in the similar situation of V4.
Positions of all new variable stars are given in
Table~\ref{pos}, and finding charts in the Fig.~\ref{find}

After the appearance of the latest version of
The Catalogue some suspected new variables were
reported in S02. We scanned our images to confirm their discoveries,
but it was not successful in any cases.

\begin{table}
\caption{Errata of the positions of \citet{Cl},
position of new variables and non-variable stars found in the
instability strip}
\centering
\begin{tabular}{@{}lccl}
\hline
ID & $\alpha(2000)$ & $\delta(2000)$ & type \\
   & [h:m:s] & [$\circ:\prime:\prime\prime$] &  \\
\hline
V238  & 13:42:15.75 & 28:18:17.2 & EW \\
V239  & 13:42:09.87 & 28:22:15.9 & RRab \\
V240  & 13:42:09.46 & 28:22:35.3 & RRc \\
\hline
N1  = NV291   & 13:42:18.06 & 28:22:40.1 & SXPhe \\
N2    & 13:42:12.41 & 28:22:34.4 & RRc \\
N3  = NV292  & 13:42:11.17 & 28:21:54.1 & RRc \\
V250n & 13:42:10.48 & 28:22:52.9 & RRab   \\
V250s & 13:42:10.54 & 28:22:52.1 & RRab   \\
V270n & 13:42:11.95 & 28:23:32.7 & RRab   \\
V270s & 13:42:11.96 & 28:23:31.9 & RRab   \\
\hline
vZ525 & 13:42:07.70 & 28:21:52.9 &  \\
      & 13:42:12.64 & 28:22:11.4 &  \\
      & 13:42:01.85 & 28:22:36.1 &  \\
      & 13:42:10.46 & 28:22:47.4 &  \\
\hline
\end{tabular}
\label{pos}
\end{table}

\begin{figure*}
\includegraphics[width=19cm]{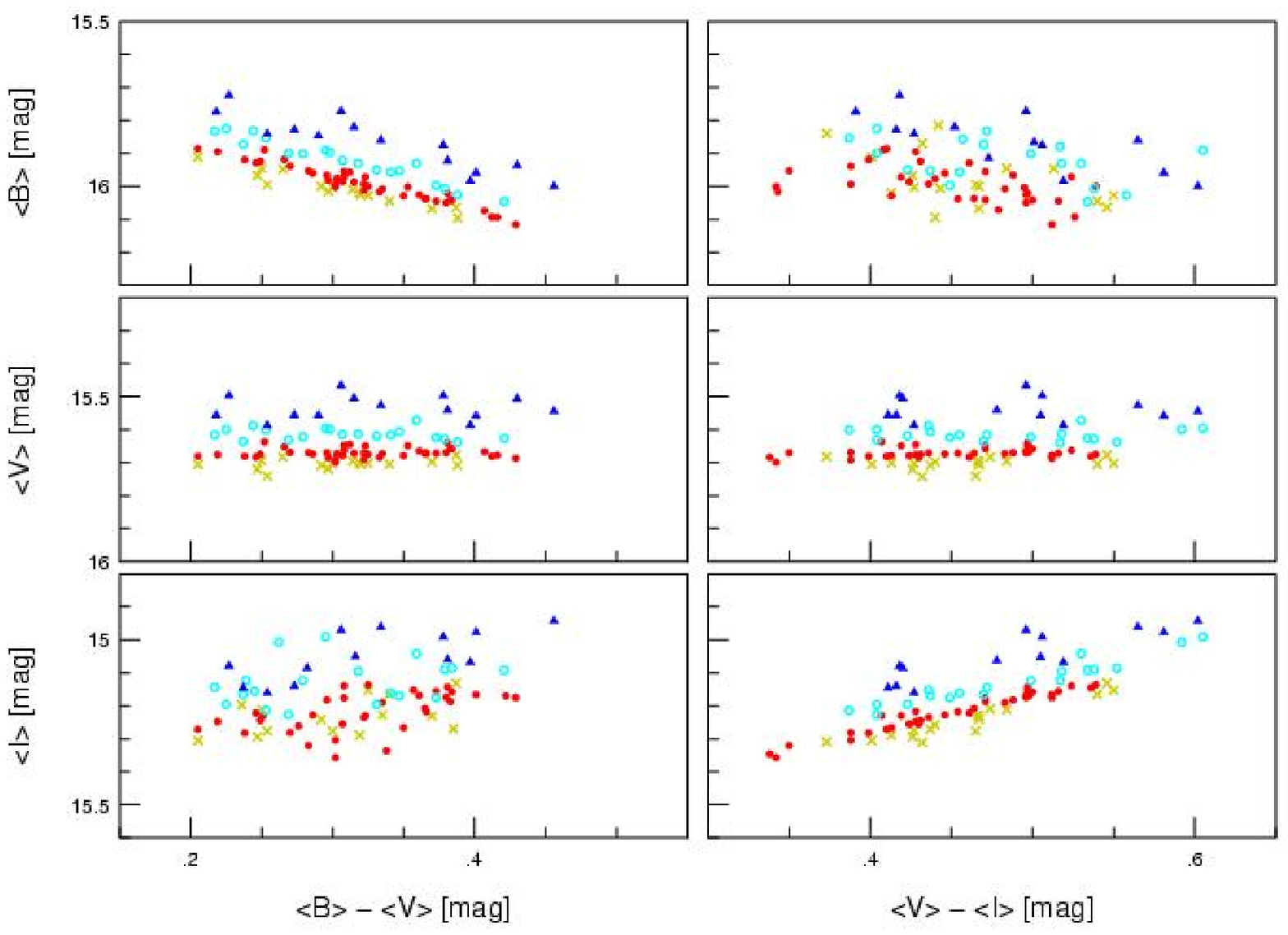}
\caption[]{Correlations among colours
 and color-indices of single mode RRab stars in M3.
Yellow times crosses, red dots, blue open squares and filled blue
squares denote the same luminosity/evolutionary subgroups as
separated by \citet{letter} among RRab variables.}
\label{hrd}
\end{figure*}

\subsection{Single mode RR Lyrae stars}

\begin{figure}
\includegraphics[width=9cm]{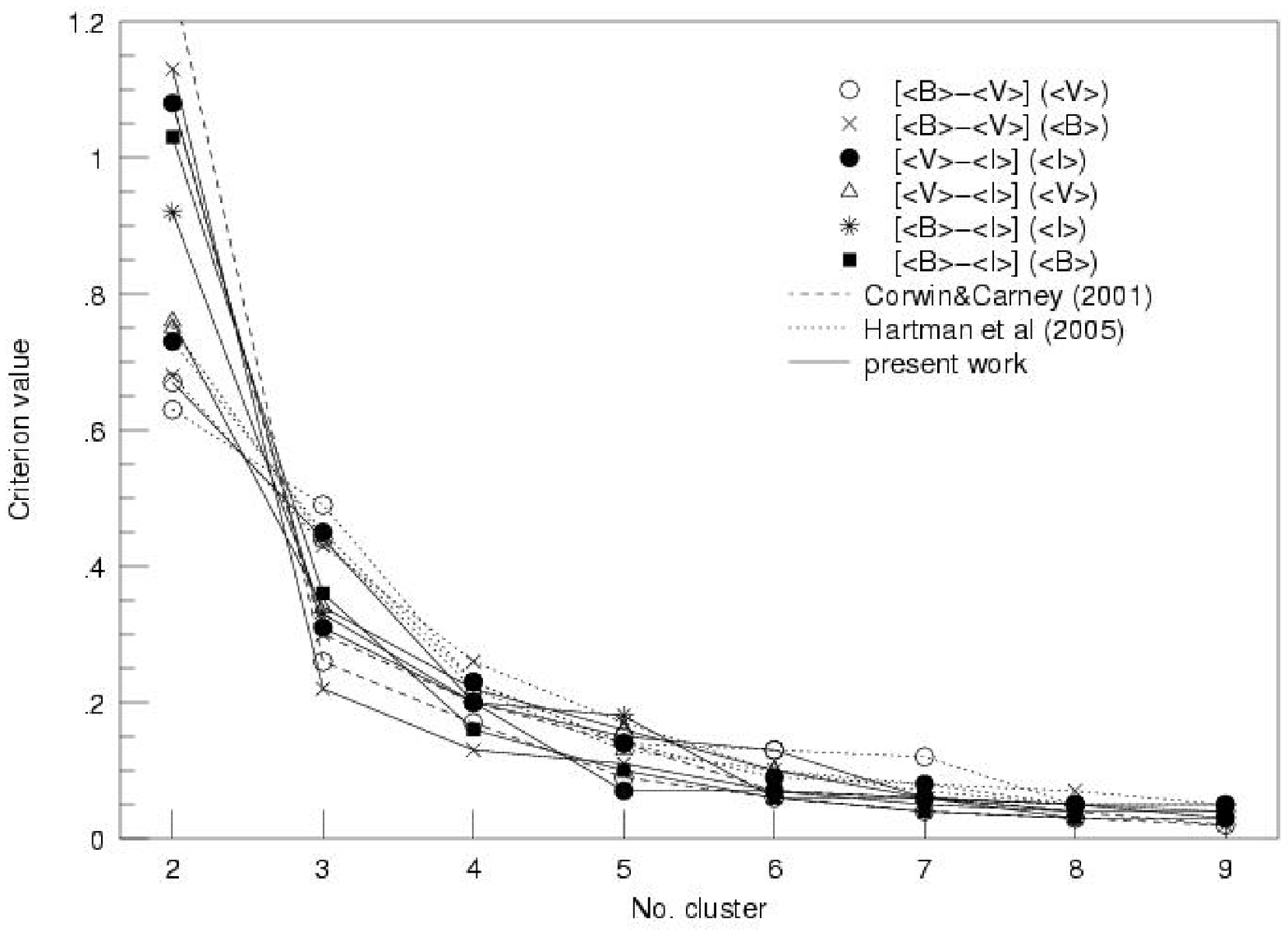}
\vspace*{-0.5cm}
\caption[]{
Criterion value in the function of the number of clusters determined by
cluster analysis of CMDs.  Different observational sets and CMDs are
encoded by line and dot styles mentioned in the figure.  The cut-off
points of functions suggest three or four intensity groups.}
\label{part}
\end{figure}

\begin{table}
\caption{Connections between the found {\it clusters\/}
by cluster analysis
and {\it intensity groups\/}
defined by \citet{letter} in the CMDs
assuming different number of groups/clusters.}
\centering
\begin{tabular}{@{}*{5}{l}}
\hline
group ID$^1$ & \multicolumn{3}{c}{No.~stars} & CMD type\\
             &  group$^2$ & cluster &  common &  \\
\hline
 1+2+3       & 76  & 76 &  75    & [$\langle B \rangle - \langle V
\rangle] (\langle V \rangle)$  \\
 1+2         & 57  & 59 &  55    &  \\
 3           & 19  & 17 &  14    &  \\
 4           & 14  & 14 &  13    &  \\
\hline
 1+2+3       & 73  & 72 &  69     &  [$\langle V \rangle - \langle I
\rangle] (\langle V \rangle)$ \\
 1+2         & 53  & 54 &  48     &  \\
 3           & 20  & 18 &  10     &  \\
 4           & 13  & 14 &  10     &  \\
\hline
 1+2+3       & 73  & 62 &  57     & [$\langle V \rangle - \langle I
\rangle] (\langle I \rangle)$ \\
 1+2         & 53  & 49 &  47     &  \\
 3           & 20  & 13 &  10     &  \\
 4           & 13  & 24 &  8      &  \\
\hline
 1+2+3       & 76  & 67 &  51     & [$\langle B \rangle - \langle V
\rangle] (\langle B \rangle)$ \\
 1+2         & 57  & 54 &  42     &  \\
 3           & 19  & 13 &  9     &  \\
 4           & 14  & 23 &  8     &  \\
\hline
\end{tabular}
\footnotesize{
\flushleft
$^1$The members of the groups are signed by these ID-s  in
Table~\ref{table4}.

$^2$The number of these stars are differ in 1-2 from the number of stars
given by \citet{letter}, because we used only those stars which
have good phase coverage from our observations themselves.
}
\label{cros}
\end{table}

\subsubsection{Mean magnitudes}

As \citet{Marconi} have shown the
intensity-averaged mean magnitudes of RR Lyrae stars are approximated well
by the physically relevant static brightness. The difference between
static and mean magnitudes depends on the amplitude but the maximum
value of this correction would be only $\sim 0.02$ mag for the highest
amplitude variable star in our sample of M3 (V148 at $A_{\it V}=1.38$
mag). Hence, throughout this section we investigated intensity-averaged
mean magnitudes without corrections.

\citet*{Cacciari} (CC05) have compared the
intensity-averaged magnitudes among data from CC01, C98 and K98
in details.  Observations in
{\it B\/} and {\it V\/} bands from CC01 data were found significantly
brighter than from C98 and a slightly brighter than {\it V\/}
data from K98. We found our
$\langle V \rangle$ and $\langle B \rangle$ magnitudes of common 92
non-modulated RR Lyrae stars to be $0.018$ and $0.013$
mag fainter then those of CC01. As CC05 pointed out
the zero point of $\langle I \rangle$ magnitudes of C98 is too faint.
On the basis of 46 RR Lyrae stars in common with C98 data sets the zero
point shift between our and C98 data is $-0.044$ mag while
CC05 estimated the zero point shift between standard
magnitude scale and C98 one's from the reddening calibration formula as
$-0.08$ mag. We performed a similar comparison between H05's data
and ours. Data from H05 proved to be brighter than ours for bands
$\langle V \rangle$ and $\langle I \rangle$ with
$0.017$ and $0.016$ mag and $0.008$ mag fainter in band
$\langle B \rangle$.

These differences are certainly related to the long standing problems of
absolute calibration.

\subsubsection{Colour \--- magnitude diagrams}\label{cmdsec}

As it is well known the colour \--- magnitude diagram (CMD) of the
horizontal branch of globular clusters
shows, in general, a good separation between variable and non-variable
stars.  Hence, we have
investigated the instability strip of the horizontal branch in
the [$\langle B \rangle - \langle V \rangle$] ($\langle V \rangle$) and
[$\langle V \rangle - \langle I \rangle$]($\langle I \rangle$) CMDs
simultaneously to check the completeness of variables' list. However, no
additional variable stars have been found, but four non-variable stars
seem to be located in the instability strip.

To decide whether these stars are members of the cluster or not, the
literature (\citealt{Cud}, \citealt*{Tu}) was looked through to get
proper motions of these stars but unfortunately, no measurements
have been found.  These stars are most likely in the foreground. But if
any of them proved to be a member of the cluster, it would be a serious
challenge to the pulsation theory.  The position of these
problematic stars is given in Table~\ref{pos}.  CC01 also reported
eight constant stars in the instability strip of their {\it B$-$V},
{\it V\/} CMD, but without position, therefore, we could not
compare them with our ones.

The separating lines between the
$\langle B \rangle - \langle V \rangle$ colour for
variable and non-variable stars are $\sim0.16$ mag at the blue edge of the
instability strip and $\sim0.43$ mag at the red edge. The dividing line
between the RRab and RRc variables is about $0.21$ mag although this
separation is less definite than in magnitude-averaged magnitudes. The
borders of the instability strip in the [$\langle V \rangle - \langle I
\rangle$]($\langle V \rangle$) CMD are $0.25$ mag and $0.55$ mag.
While the dividing line between fundamental and overtone pulsators is
about $0.35$ mag.
Comparing these values with previous studies we found
that both the blue and red edges of
 our instability strip are slightly bluer than
from data of H05, but similar to C98, CC01.
The width of instability strips are, however,
the same ($\sim0.27$ mag) for all three cases.

\citet{letter} have studied relations
in $\langle V \rangle$ mean magnitudes
and light curve parameters (viz. period and Fourier coefficients).
Four subgroups were separated from non-modulated RRab stars according
to their brightness and light curve shapes and these groups were
interpreted as different stages of horizontal branch stellar
evolution. The brightest group shows Oosterhoff~II properties as
opposed to the other three ones. This result suggests that the Oosterhoff
dichotomy of the Galactic globular clusters is to be an evolutionary
effect and M3 is not a typical Oosterhoff~I cluster.
In Table~\ref{table4} these groups of RRab stars are
marked with number 1-4 from the faintest group to the brightest one.
Recently CC05 have questioned the existence of these
four groups. They found a bimodal structure in the magnitude
distribution: a brighter group at $\langle V \rangle=15.52\pm0.02$
(which is in the same position as the brightest group of
\citealt{letter}), and a main body at $\langle V \rangle=15.64\pm0.04$.

Through this subsection we have reanalysed those non-modulated
RR Lyrae stars which were selected by \citet{letter}
on the basis of their good quality light curves.
The $[\langle B \rangle - \langle V  \rangle] (\langle V \rangle) $
and $[\langle V \rangle - \langle I \rangle] (\langle V \rangle) $
colour magnitude diagrams in Fig.~\ref{hrd} correspond to fig. 4 in
the paper \citet{letter}.
The similar structure of figures can be clearly recognized:
the four subgroups of RRab stars with different brightness are lying in
parallel strips, and the averaged periods are shifted from shorter to
longer according to the brightness changes from the fainter to the
brighter.
The same structures appeared in all other
diagrams in Fig.~\ref{hrd}.
Naturally, the parallel strips have different steepness
for various colours. The
groups in [$\langle B \rangle-\langle V \rangle$]
($\langle I\rangle$) and [$\langle V \rangle-\langle I \rangle$]
($\langle B\rangle$) CMDs are less evident compared to the other ones
caused most probably by higher scatter resulting
of measurements in the three independent bands.

We have examined all colour \--- magnitude relations with the
cluster analysis, an
efficient multivariate statistical method, to point out the solidity of
identification of these groups.
(We refer here to the book of \citet{Murtagh} as an excellent introduction
to the method.) The Ward's minimum variance method was chosen
from the many different clustering algorithms which is generally a very
robust process.  (It should be mentioned,
that the average linkage method has also fit
well for the data and yielded very similar results.) The algorithm is a
hierarchical type clustering which does not assume any prior
knowledge about the number of clusters. Finding the right number of
clusters is based on a horizontal ``line draw'' through the resultant
hierarchy represented by a dendrogram. The choice of where to draw the
line is derived from large changes in the criterion value.

Fig.~\ref{part} shows the criterion value \--- cluster number functions
obtained from different standardized data sets where
in addition to our data
CMDs were also used from CC01 and H05 of stars in common with our sample.
Most functions belonging to different data sets have a well defined
cut-off point where the criterion value dropped.
The $[\langle B \rangle-\langle V \rangle]$($\langle V \rangle$) ratios
derived from CMDs, based on H05 and our data set, suggest 
{\it four clusters}.
Other CMDs obtained from CC01 and present
observations seem to segmented into {\it three clusters}. 
The cut-offs are less specific at the functions of H05 data 
because of the small number of common stars (49-59 depending on colours).

\begin{figure}
\includegraphics[width=9cm]{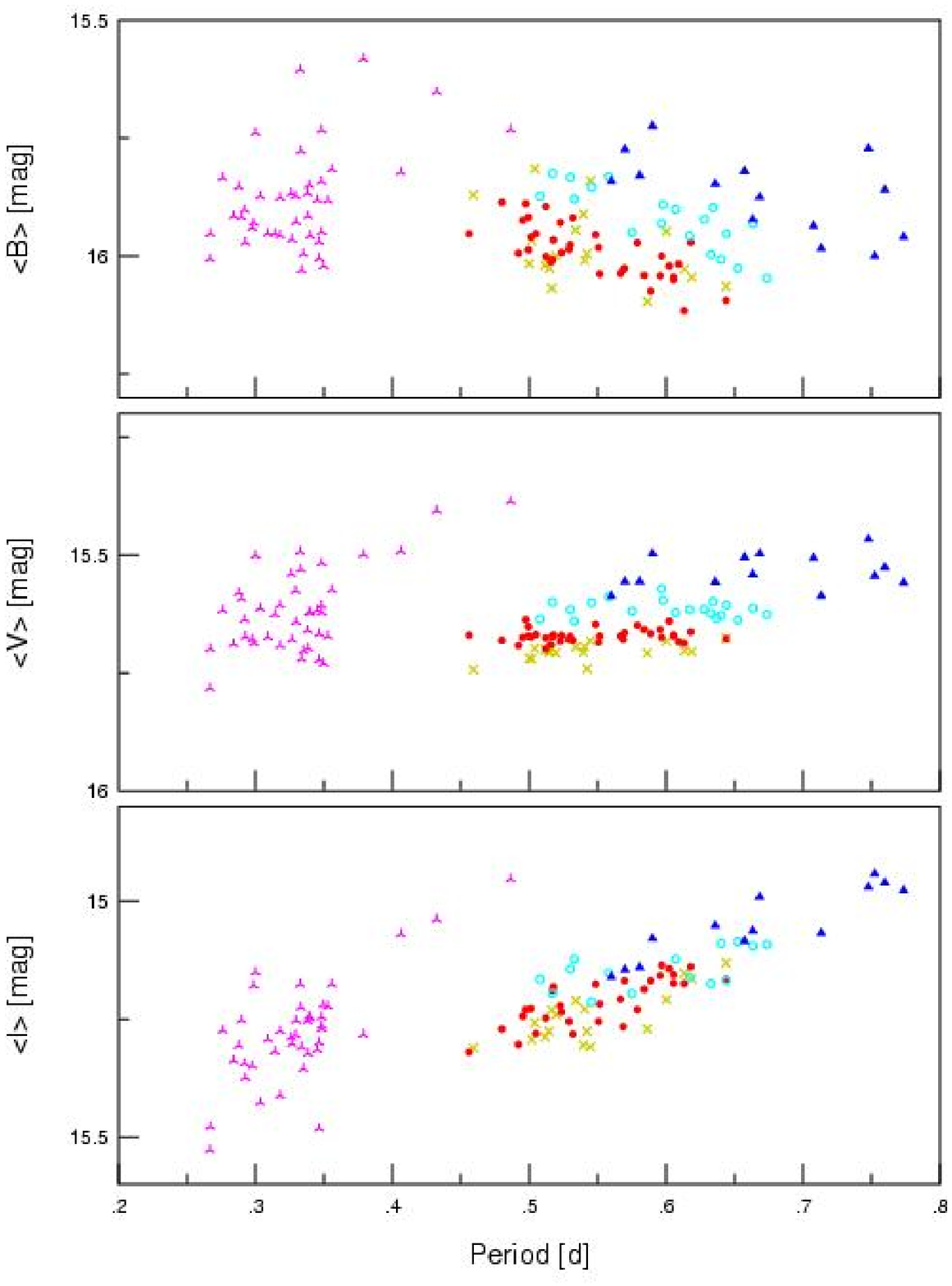}
\vspace*{-0.5cm}
\caption[]{Intensity-averaged mean magnitudes of non-modulated RR Lyrae
stars vs.~period. Symbols as in Fig.~\ref{hrd},
the purple three-prong crosses denote RRc variable stars.}
\label{pl}
\end{figure}

\begin{figure}
\includegraphics[width=9cm]{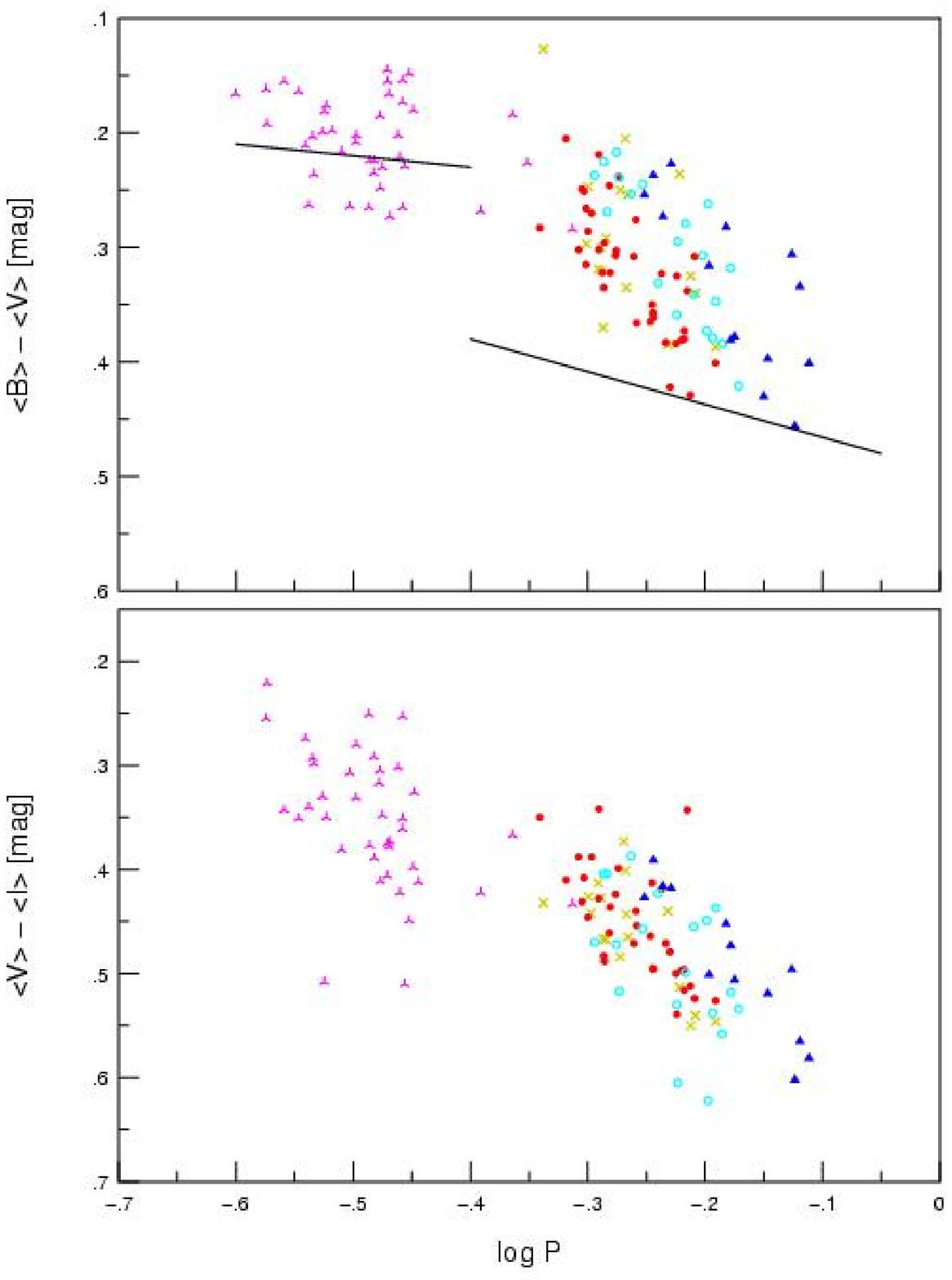}
\vspace*{-0.5cm}
\caption[]{Colour indices vs.~period. The lines indicate the predicted
edges of instability strip taken from \citet{Marconi}
at the mixing length parameter of 1.5.
Other symbols are as in Fig.~\ref{pl}}
\label{pi}
\end{figure}

To investigate the connection of signed groups in  Figures~\ref{hrd},
\ref{pl} \--- \ref{pi} and groups found by the cluster analysis the
star content of each cluster has been checked and the result are given
in Table~\ref{cros}. 
{\it a)\/} When we assumed two subgroups according to 
CC05 and groups signed with 1, 2 and 3 by \citet{letter}  
were handled together while group~4 
remained separated the two clusters obtained from cluster analysis 
corresponded well to these sets both in number of elements 
and star content.  In the case of   
$[\langle B \rangle-\langle V \rangle]$($\langle V \rangle$)
CMD both analyses yielded 76 and 14 
members clusters with 75 and 13 overlapping of star contents, respectively.
Similar connections can be realised for the $[\langle V \rangle-\langle I 
\rangle]$($\langle V \rangle$)  CMD but less overlapping star contents.
Using elementary formulae of statistics the probabilities of such a 
connections are $p=1.17\cdot10^{-13}$ and $p=4.8\cdot10^{-12}$, 
respectively. Overlapping structures for the other two CMDs are not
statistically significant.
{\it b)\/} However, our cluster analysis suggested at least three clusters 
(Fig.~\ref{part}). In the models with three sets (groups 1+2, 3 and 4 vs. 
three clusers) 
$[\langle B \rangle-\langle V \rangle]$($\langle V \rangle$) 
and $[\langle V \rangle-\langle I
\rangle]$($\langle V \rangle$) CMDs yielded again satisfactory result. 
The probability that these structural similarities are only incidental
is very low $p=6.2\cdot10^{-15}$ and $p=3.4\cdot10^{-14}$. {\it c)\/}
No significant connections were found between four groups and 
four clusters models. 

As a conclusion of this section we say that our cluster analysis
on CMDs found the same two sets of stars as they were found on 
the basis of
Bailey diagram and period distribution by CC05 and the same three 
sets which were identified by Fourier parameters in \citet{letter}.

\subsubsection{Period \--- brightness and period \--- colour relations}
\label{pbc}

In the panels of Figure~\ref{pl} the magnitudes vs.~period are plotted
where the four above discussed luminosity groups  are
indicated with the same
symbols as in Figure~\ref{hrd}.
The highest scatter of plot
$P(\langle B \rangle)$ is originated  probably not
only on account of the higher noise level of observations, but
due to the fact that the effects of luminosity and temperature
variations upon the expected periods are almost orthogonal in this
plane (see \citealt*{Catelan}).

In the panels of Fig.~\ref{pl} RRab and RRc stars are well
separated, except for four variables: V70, V129, V170 and V261.  Most
likely, these variables belong to a distinct group on the basis of their
common properties. They have asymmetric light curves with amplitude
between RRab and RRc variables. Their periods are also between
those of the two
groups and mean brightnesses for all filters are higher than the other
RR Lyrae stars at a given period.  V70, V85, V129, V170 and V177 have
been characterized by CC05 as ``long P / overluminous''
stars. H05 and present data show V85 and V177 to be regular RRc stars.

A definite correlation can be seen between period and magnitude
$\langle I \rangle$ of RRab stars. Applying a linear least-square fit
for 76 non-modulated RRab variables, we have found
\begin{equation}\label{PL}
\langle I \rangle = -1.453 \log P + 14.828,
\end{equation}
with $rms=0.04$ mag. To our best knowledge the work of \citet{Layden}
was the first paper in which a period-luminosity type relation was derived
in the $I$ band for the RR Lyrae stars in NGC3201. \citet{Pritzl}
discussed this relation in the case of the globular cluster NGC6441.
\citet{Catelan} demonstrated that this type of relations must exist.
They plotted period\---luminosity type graphs in standard $UBVRIJHK$
photometric bands, for some different metallicities and horizontal
branch morphologies (evolutionary states) based on synthetic horizontal
branch calculations. Fig.~\ref{hrd} is qualitatively similar to figs 6-9
in the paper of \citet{Catelan}. In addition,
we derived the period\---luminosity relation for M3
with the formula given by  \citet{Catelan}
accepting the parameters Lee-Zinn
type horizontal morphology index and metallicity as
0.282 and $Z=0.001$, respectively.
The result is $M_I=-1.644\log P -0.234$ which is also close to the
observed relation Eq.~(\ref{PL}).

We have to mention that no clear correlations can be identified for RRc
variable stars in contrast to RRab ones. 
However, converting the periods of RRc variables to their fundamental
mode equivalents ($P_{\rm f}$) RRc stars fit well to the 
period -- $\langle I \rangle$ relationship extending it to shorter
periods. What is more, fundamentalized periods of overtone pulsators
define the same relation within the fitting error ($rms=0.07$ mag) 
as Eq.~(\ref{PL}): $\langle I \rangle = -1.486 \log P_{\rm f} + 14.748$. 
The range of period for the RRc stars is smaller than for RRab, so 
the relation between period and $\langle I \rangle$ is  
naturally less evident and
noisier than from RRab stars but it is even so significant.

In  Figure~\ref{pi} the log P \--- colour relations are plotted.
To harmonize CC01 data with the theoretical
models \citet{Marconi} assumed an increasing in the
mixing-length parameter from 1.5 to 2.0 from the blue edge to the red
one. In the upper panel of the Fig.~\ref{pi} the predicted edges of the
instability strip at the mixing-length parameter $1/H_{\rm p}=1.5$
(taken from the paper \citealt{Marconi}) are also shown.  On the basis
of our data also an increasing mixing-length parameter seems to be
adequate but varying form $\sim 1.1$
to $\sim 1.5$. Considering interstellar reddening the
situation does not change much because the reddening is
$E(B-V)=0.01\pm0.01$ for M3 obtained by different methods (for a review
see CC05).

\subsection{Variable stars with Blazhko effects}

The amplitude and/or phase modulation of several RR Lyrae stars was
discovered by \citet{Blazhko}, but the correct explanation of the
effect is still missing (see \citealt{Dziem} for present situation).
Since M3 contains the highest number ($\sim70$) of RR Lyrae stars in one
cluster showing Blazhko effect, this offers a unique opportunity to
investigate this interesting phenomenon. Unfortunately, the typical
accuracy of the earlier photographic observations (see for a review and
complete references \citealt{Bela65}, 1973, \citealt{Kuk61}, 1970,
\citealt{Welty}) were inadequate to studying the Blazhko effect in
details. A basic parameter such as the period of the Blazhko cycle was
determined only for one star (V5 by \citealt{Panov}). On the other
hand, high quality CCD observations are badly
sampled and they typically consist of about a dozen of nights spreading
over 1-3 years.

\begin{table}
\caption{Estimated modulation periods $P_ m$ of RR Lyrae
stars with Blazhko effect. The sources of used data and total number of
observed nights are also indicated. The slight phase modulations
are noted by `phase' in column remark.}
\centering
\begin{tabular}{@{}*{5}{l}}
\hline
ID & No.    & ref. & $P_m$  & remark \\
   & nights &            & [day]         &         \\
\hline
V66   & 59  &  1,2,3,4  & 56.24          &        \\
V67   & 59  &  1,2,3,4  & 80.94          &        \\
V10   & 52  &  1,2,3,4  & 124.78         &        \\
V43   & 51  &  1,2,4    & 73.54          &        \\
V78   & 51  &  1,2,4    & 38.65          &        \\
V121  & 51  &  1,2,4    & 64.37          & phase  \\
V104  & 50  &  2,3,4    & 63.78          &        \\
V59   & 43  &  2,3,4    & 62.14          &        \\
V63   & 43  &  2,3,4    & 44.91          &        \\
V101  & 41  &  2,4      & 91.15          &        \\
V110  & 41  &  1,2      & 45.66          &        \\
V7    & 40  &  2,4      & 61.9:          &        \\
V47   & 40  &  1,2      & 212.06:        & phase  \\
\hline
\end{tabular}

\footnotesize{1. \citet{C98}, 2. \citet{CC01}, 3. \citet{K98}, 4.
\citet{hartman}}
\label{Blazhko}
\end{table}


\begin{figure*}
\includegraphics[width=16cm]{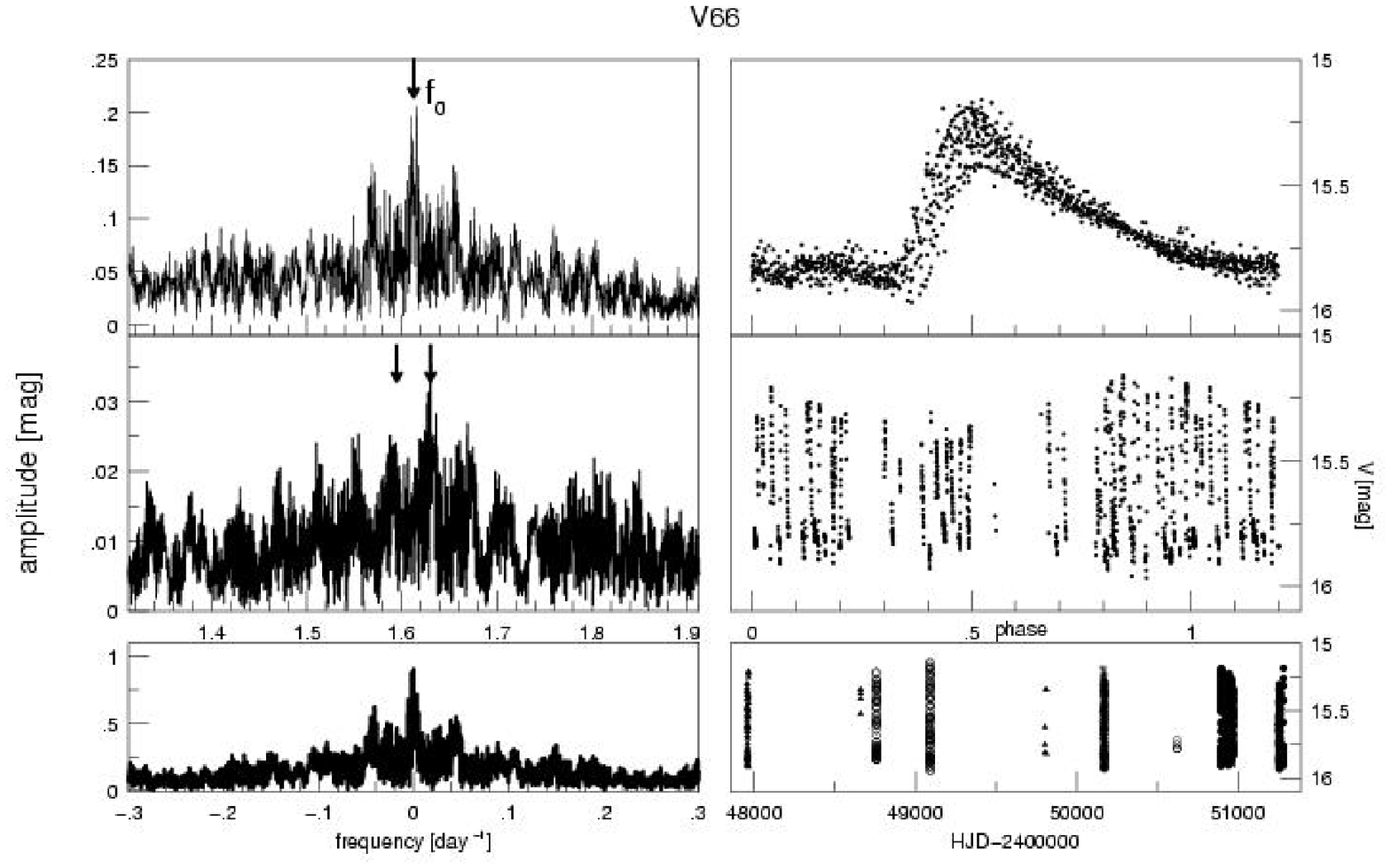}
\caption[]{
An example for Blazhko period finding. Figures for all stars in
Table~\ref{Blazhko} are available in electronic form. Combined {\it V\/}
light curve folded with the best pulsation period (top right). Fourier
amplitude spectrum of combined data around the strongest peak signed by
$f_0$ (top left). Residual spectrum of prewhitened data with $kf_0$,
$k=1,\ldots,10$, where the $f_0\pm{f_m}$ frequencies are noted by arrows
(middle left). Light curve folded with modulation period (middle
right). Data distribution (bottom right), where the symbols denote sources
viz. triangles:
C98, open circles: CC01, stars: K98, crosses: H05, dots: this work.
The spectral window is shifted to $f_0$ in the same scale as the above
amplitude spectra (bottom left).}
\label{v066lcsp}
\end{figure*}

Studies based on the large surveys
(MACHO: \citealt{Alcock2000}, 2003 and OGLE: \citealt{Moskalik})
distinguish among different type
of modulations based on the structure of the Fourier spectrum of stars.
As Jurcsik et al. (2005a,b) have expressed this classification likely
resulted in too many subtypes. The classical BL-type (frequency triplet
with two equidistant, side lobes around the main frequency) and type
$\nu_1$ (frequency doublet) may differ only in their data distributions:
sampling, S/N ratio. Since the number and time coverage of our data are
rather limited these subtypes were not separated and both are called
as Blazhko modulation.

The light curves of all observed RR Lyrae stars
were searched for modulation. Revising
the literature from point of view of Blazhko behaviour some controversial
cases were found.  In the cases of  V10, V22, V54, V44, V140 and V218
the Blazhko modulation was verified. In the case of V18 we cannot
detect any light curve variations on the basis of all CCD observations
in contrast to earlier photographic studies. In addition, we found evident
light curve changes for V104, V168, V172, V191 and V211 for
the first time.  Details of individual stars are discussed in
Sec.~\ref{ind}.

Our technique for obtaining average brightnesses
of Blazhko stars is described in
Sec.~\ref{Avbr}. The distributions of Blazhko type RR Lyrae in the
period \--- luminosity diagrams are similar to the non-modulated
variables, but their numbers are significantly less at longer periods
(higher brightnesses). \citet{Alcock} found a similar effect in
the case of LMC variables and CC05 demonstrated
it in the case of M3 as well.

\subsubsection{Search for Blazhko period}
\label{pbcBl}

The period of Blazhko cycle is an important parameter of these stars.
Since the typical modulation period is about 10-500 days
\citep{Smith}, it cannot be determined from a data set optimized
to pulsation cycles of RR Lyrae stars.
To determine Blazhko period all CCD
{\it V\/} data of all modulated RR Lyrae stars in the cluster
were collected. In preparing homogeneous data sets we had to
take into account the
zero point shifts among different photometries which differ from star to
star. \citet{RRGem} have demonstrated that the light curve changes are
concentrated in a narrow phase interval ($\sim0.2$) of pulsation: 0.4
\--- 0.6, if maximum is at 0.5. We confirmed this result for all RR Lyrae
stars in M3 with pure amplitude modulation. The residual phase diagrams
folded with pulsation period after prewhitening with the mean pulsation
light curves clearly showed the narrow range of variability.  Strong phase
modulation, however, removed the effect.

Applying this result we prepared
phase diagrams for each Blazhko type star from each pair of data sets
containing the present and one of the published data sets.
Phase diagrams were divided into 10-15 bins, then
the  {\it average of standard deviations of the bins\/} (ASDB)
was calculated so that bins around the maximum phase were omitted.
We searched for the {\it minimum\/} of the ASDB while data set
taken from the literature was shifted by 0.01 mag at each iteration.
The shifting value at the minimum of the ASDB yielded the
proper zero point to combine a given pair of data sets.
The process was carried out for each star and pair of data sets.

Figure~\ref{v066lcsp} demonstrate the process of modulation
period search. The spectra were computed by discrete Fourier
algorithm included in {\sc mufran} program package by \citet{Zoli}.
The combined data were Fourier analyzed to refine the pulsation frequency.
The prewhitened spectrum (with the pulsation frequency and its
harmonics) was searched to find close frequenci(es) to the main one.
To verify the estimated modulation frequency window spectrum was
compared with the spectrum of data and light curve folded with modulation
period was also prepared.

The results of this period search are shown in Table~\ref{Blazhko}. For
13 variable stars a more or less established value of modulation period
has been found for the first time.  Length, distribution and/or quality
of data of other stars were not sufficient for finding a modulation
period.  Let us mention here that just four stars (V10, V34, V66, V67)
are included in all published CCD surveys (C98, K98, CC01, H05 and present
work)!  Only 37 stars were observed at least in 40 nights by CCD which
amount of observations proved to be necessary to estimate a
Blazhko period.  The found periods are within the normal range of
Blazhko cycles and far below the borderline of limit periods
discovered by \citet{HannaActa}.

Some words about modulated overtone pulsators V140 and V168. 
There were found frequency triplets/doublets at numerous RRc stars
in different globular clusters
(Olech et al. 1999a,b, 2001, \citealt{Walker}, \citealt{CR}), but
no previous studies reported modulated RRc stars in the case of M3. As it 
was stressed above combining of different photometries works properly
only for pure amplitude modulation. V140 and V168 have strong phase
modulations and their amplitude changes are only marginal so
their different photometric data sets
had to be analyzed separately. In the Fourier spectrum of prewhitened
data of V140 two peaks and their harmonics appeared equidistantly
from the main frequency and its harmonics respectively,
indicating a modulation
period about 9 \--- 14 days. However, consistent period could not be
found. In contrast to V140 the residual spectrum of V168 shows only one
peak and its harmonics at $f_1=3.4754158$ day$^{-1}$. If we interpret
it as $f_1=f_0-f_m$ modulation frequency the Blazhko cycle would be
6.743 days.  An alternative explanation for the peak is excitation of a
non-radial mode as it was discussed by Olech et al. (1999a,b).

\subsection{Double mode RR Lyrae stars}

Presently 74 galactic globular clusters are known containing 
RR Lyrae stars \citep{Cl}. 
 About 3 per cent of the total number of $\sim2000$ 
cluster RR Lyrae stars are double mode (RRd) pulsators. Their 
distribution among clusters is, however, highly non-uniform:
only six clusters (NGC2419, M68, M3, IC4499, NGC6426 and M15) 
contain at least one RRd stars. The richest clusters in double mode
RR Lyrae stars are M68 (12 RRd, 29 per cent), M15 (17 RRd, 19 per cent)
 and IC4499 (17 RRd, 18 per cent). The rate of RRd stars of M3 (9 
RRd, 3 per cent) does not exceed the averaged rate. But it is 
still noteworthy since we know clusters overall rich in RR Lyrae stars 
but no RRd stars were detected in them (e.g. $\omega$~Cen, M62, M5).

A series of studies (\citealt{NC89}, \citealt{Clement97},
\citealt*{CCRRd}, \citealt{CC04} (CC04), and further
references therein) revealed double mode RR Lyrae content of M3.
In these studies eight stars were
identified as RRd variable: V13, V68, V79, V87, V99, V166, V200 and
V251. \citet{hartman} have suggested additional four double mode
candidates: V252, V253, V270 and NV290. They stressed that the
multi-periodic behaviour of these stars were determined by eye without
systematic search so we have investigated light curves and Fourier
spectra of the first three candidates. V252 proved to be an RRd
variable indeed (see the following section). As it was mentioned
before, the strange brightness variation of V270 is caused by two
overlapping RR Lyrae stars and V253 seems to be a normal RRc star.

\subsubsection{V252 as a new RRd star}\label{V252}

\begin{figure*}
\includegraphics[width=18.5cm]{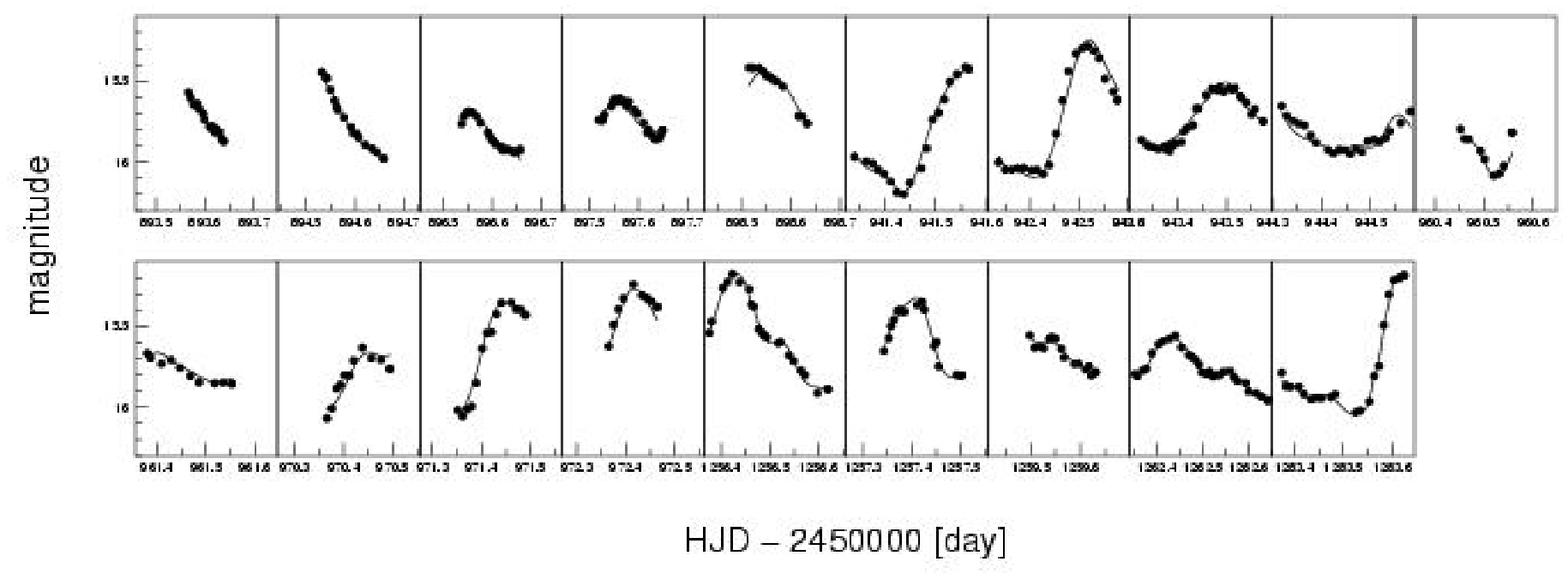}
\caption[]{Observed and fitted {\it V\/} light curve of the new RRd
variable V252. Each night was plotted on the same size of time interval.}
\label{lc252}
\end{figure*}

\begin{figure}
\includegraphics[width=8.8cm]{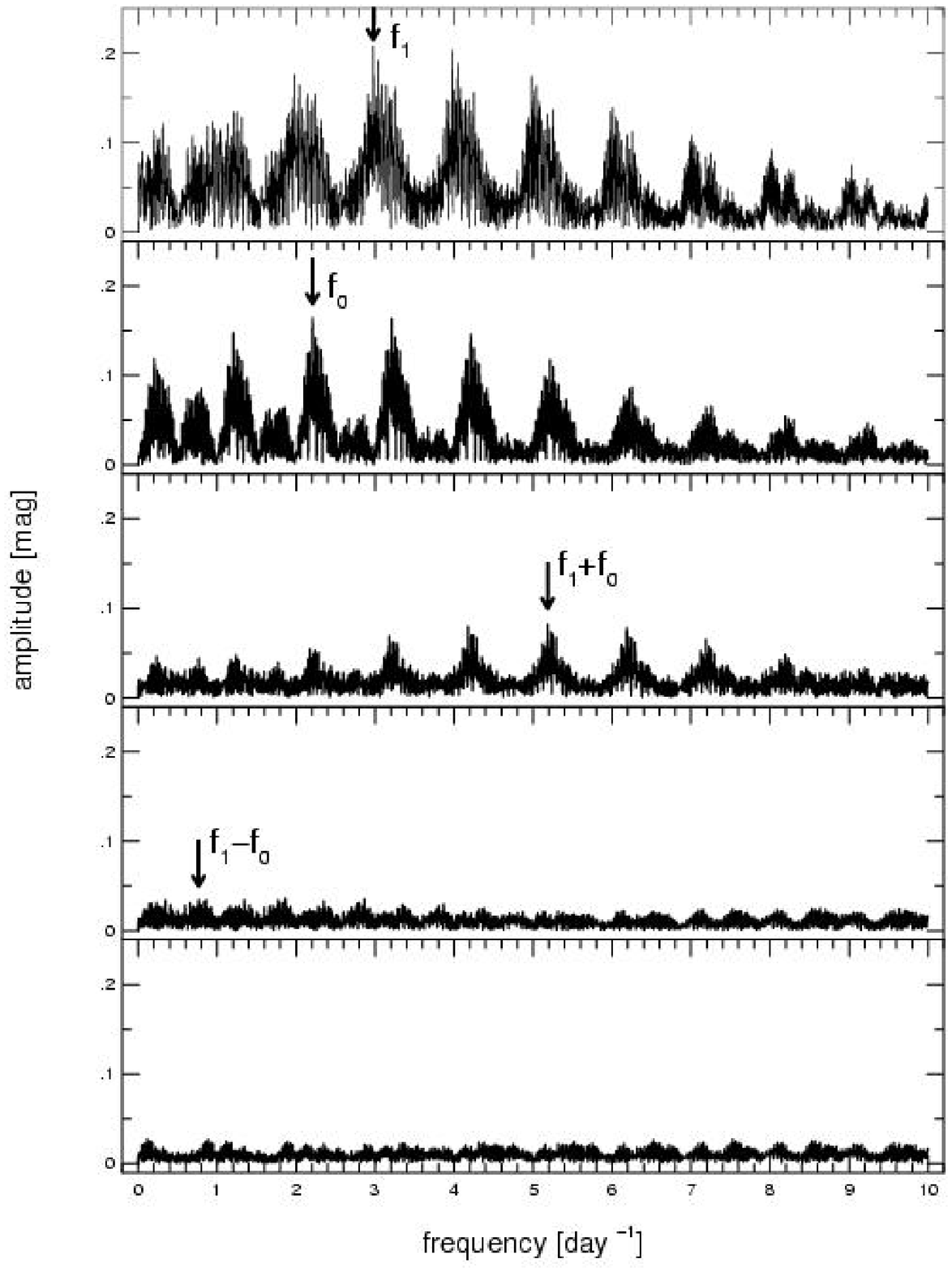}
\caption[]{Fourier amplitude spectra for the RRd variable V252.
The frequency searching is demonstrated
step by step from top to bottom panels.}
\label{sp252}
\end{figure}

From our light curves it turned out that V252 shows double mode
behaviour. The uppermost panel in Fig.~\ref{sp252} presents the
amplitude spectrum of the {\it V\/} data.  The next panel shows the
spectrum after
prewhitening with the dominant frequency $f_1=2.97628$ day$^{-1}$ and
its six harmonics.  The third panel is a plot of the prewhitened
spectrum with $f_0=2.208528$ day$^{-1}$ and its seven harmonics. In
this spectrum the linear combinations of the two radial modes $f_1+f_0$
can clearly be identified and we get the fourth spectrum after that
we whitened out with it.
The $f_1-f_0$ linear combination is much less
significant than $f_1+f_0$ in the previous spectrum, however,
considering this odds the r.m.s.~of the least square fitted light curve
has been improved well. The lowest panel shows the final whitened
spectrum. No additional peak can be recognized.
Fig.~\ref{lc252} was obtained so that the light curve was fitted
simultaneously by all frequencies obtained from the above process. The
$1\sigma$ error of the fit is $0.04$ mag which is slightly larger than the
intrinsic observational error indicating other linear combinations
and/or higher harmonics in the data.  Because of their low amplitude it
is impossible to find them from these data.

\subsubsection{Period and period ratios}

\begin{figure}
\includegraphics[width=8.5cm]{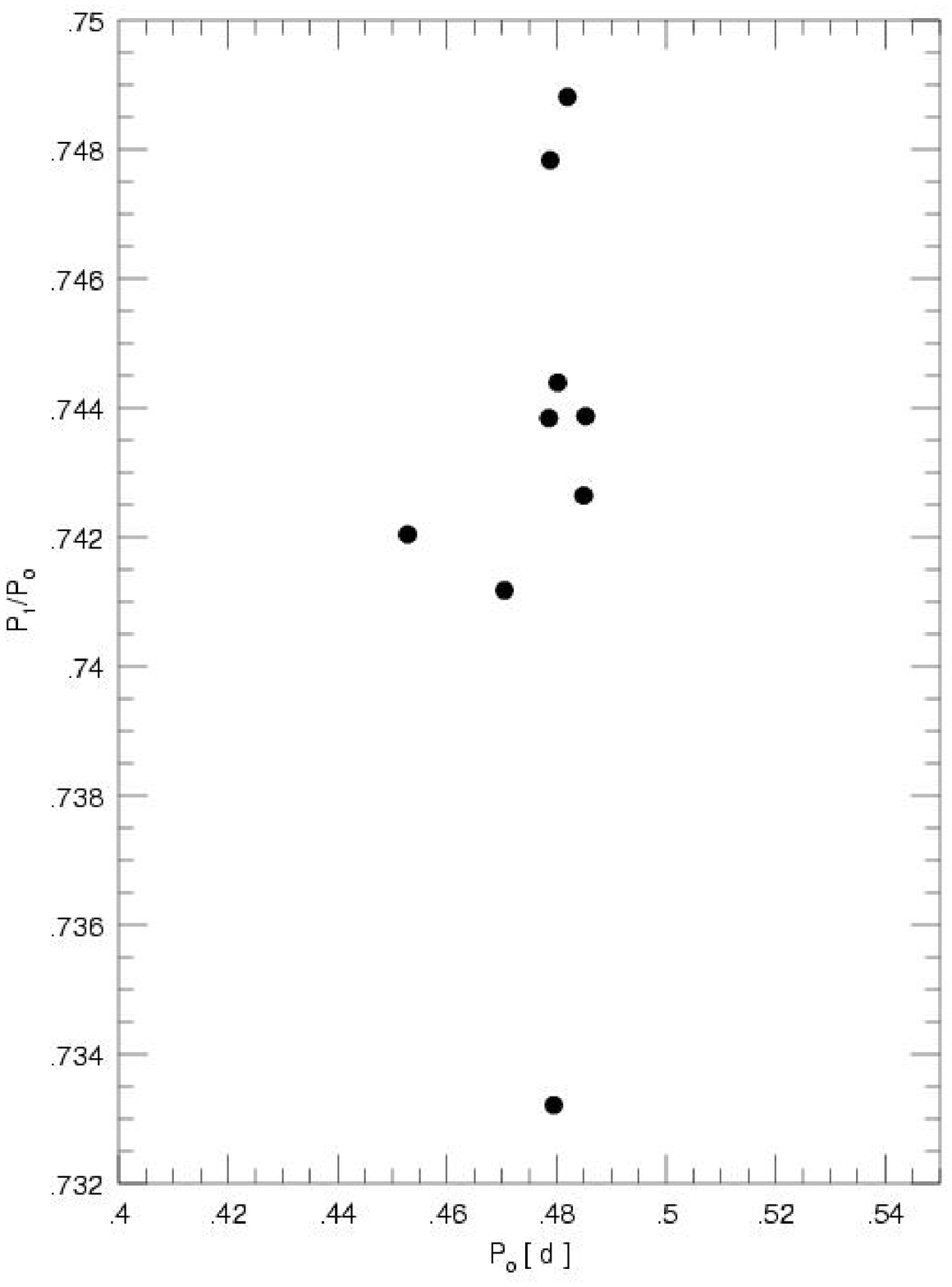}
\vspace*{-0.5cm}
\caption[]{Petersen diagram of double-mode RR Lyrae in M3.}
\label{peters}
\end{figure}

The periods of all measured double-mode variables were determined as
it was described for V252 (see Tab.~\ref{table4} for the values). The
periods determined by us are generally in good agreement with the
values given by CC04. The period of fundamental mode ($P_0$) of V79
started to decrease drastically in the middle of the last century
\citep{V79} and we found it still decreasing.
The collected data from the two recently
recognized RRd stars V200 and V251 allowed us to specify their periods.
The new periods and period ratio of V200 have placed it among the
`normal' double-mode RR Lyrae stars as opposed to the work of CC04.
However, the Petersen diagram of the cluster in Fig.~\ref{peters} shows
that V13 is very far from all other RRd stars. What is more, V13
has the smallest period ratio ($\sim0.734$) known among double mode RR
Lyrae stars. The period of V13 has been decreasing dramatically: it was
$P_0=0.4830311$ d in the 60s (calculated from the data of
\citealt{Bela65}) and it is now $P_0=0.4795043$ d.  Both facts are
alluding to strong evolutionary effects. V13 is maybe a `transient'
double-mode pulsator and not a `stable' one (for definition of
terms see \citealt{Robi}).

\subsubsection{Sudden changes in modal contents}\label{RRds}

\begin{figure*}
\includegraphics[width=18cm]{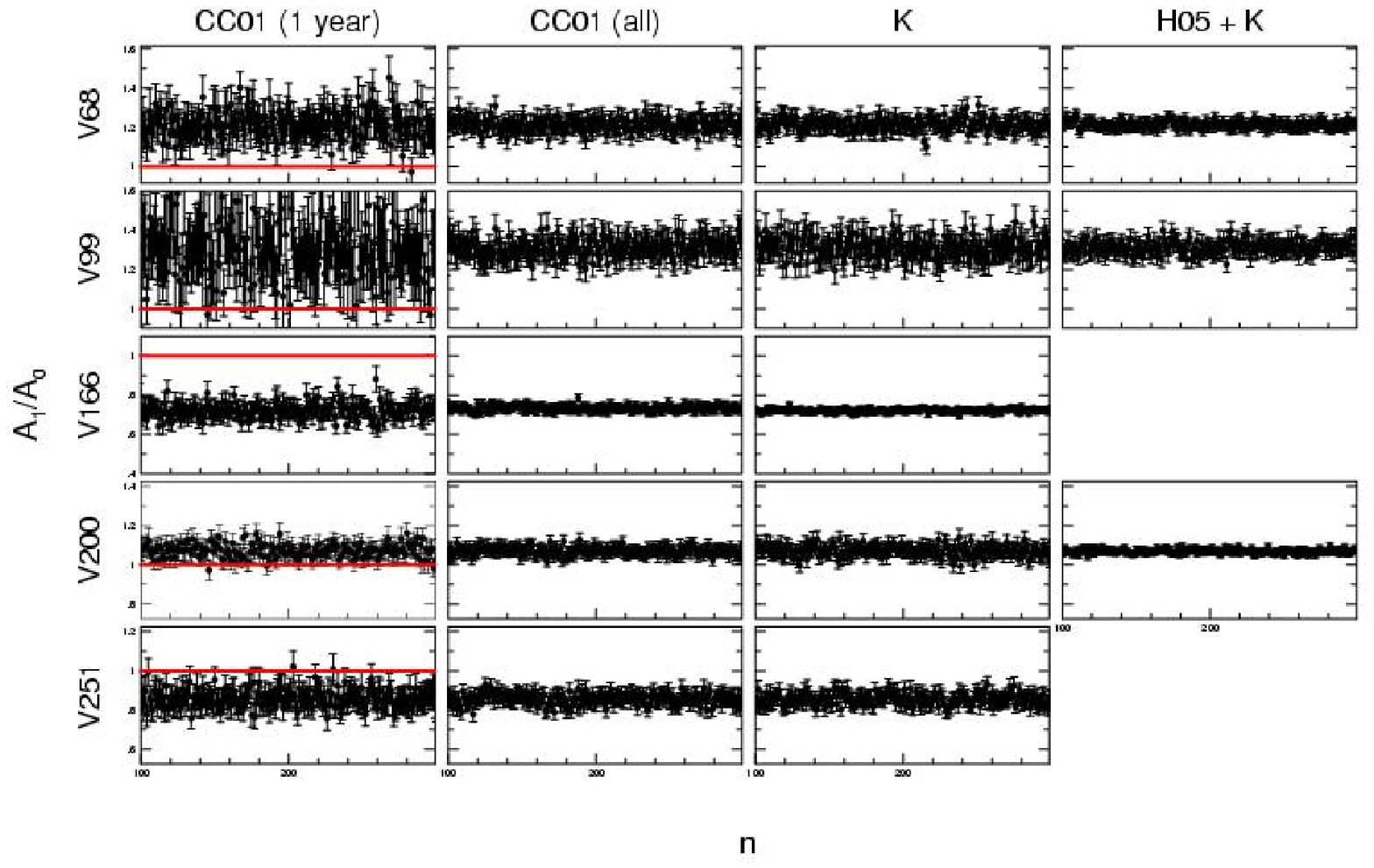}
\caption[]{
Calculated amplitude ratios ($A_1/A_0$) of synthetic light curves
versus iteration number $n$. Here $n$ are defined by $T_0=t_0+n*0.01$
days; $t_0$ and $T_0$ are the epoch of the beginning of the synthetic
light curves and a given sampled ones, respectively. The rms error of
the fit is also marked for each point. Data distributions were taken
from CC01 = \citet{CC01}, H05 = \citet{hartman}, K = present work.
}
\label{sym}
\end{figure*}

Concerning its double mode pulsators M3 seems to be unusual. The
cluster is unique among the galactic globular clusters in that it has
RRd variable stars with a dominant fundamental mode: V13, V166 and V251.
Only very few RRd variables are presently known to have a dominant
fundamental mode, and they all are field variables: AQ~Leo, NSVS 5222076 
\citep*{Oaster} and some stars in the LMC \citep{Alcock2000}.

M3 is so far the only cluster in which RRd stars have been observed
to switch from one dominant mode to another while remaining RRd type
variables.  Variations in the double-mode behaviour were observed
in a number of RRd variable stars (\citealt*{Jerzy},
\citealt{JurcsikBarlai}, \citealt*{Clement93}, 1997, \citealt{Purdue}).
These variations concern initiation or cessation of the double-mode
behaviour and changes in the length and amplitude ratios of the two
periodicities. None of these studies reports a switch between dominant
modes. However, for five RRd
stars (V68, V99, V166, V200, V251) of the known nine in the globular
cluster M3 such a possible mode switch was published by different
authors (\citealt{CCRRd}, \citealt{BJ}, CC04). The case of V79 is
different. As \citet{Clement97} revealed this star switched from a
single mode first overtone pulsation to double mode one.

It has to be stressed, that except for V79, the alleged mode switching
was found within 1-2 years long time series. The question is whether
they were real effects or just artifacts of data distributions?

To answer the question the best of (longest, homogeneous) CCD
data sets were chosen as {\it reference light curve\/} for each
RRd stars and were Fourier analyzed. First the frequency
of the dominant mode was determined, then, after prewhitening the data
with this frequency and its low order (7-15) harmonics, the frequency
of the other mode was searched. If the spectral window did not allow to
make clear distinction of the true period because of severe aliasing,
then that possible frequency was accepted which was the closest to the
frequency of a given mode as determined from other observations.

Using the found frequencies of the fundamental and first overtone
modes, their harmonics and linear combinations in each data set, least
square fit solutions were calculated in order to determine the
amplitudes. The fitting formula was
\begin{equation}
m(t)=a_{00}+\sum_{i,j} a_{ij}
\sin[2 \pi  \nu_{ij}(t-t_0)+\varphi_{ij}],
\label{fit}
\end{equation}
where the designations are the usual: the Fourier amplitudes, $a_{ij}$,
phases, $\varphi_{ij}$, the epocha, $t_0$, the frequencies,
$\nu_{ij}=if_0+jf_1$, $i$ and $j$ integers. As the parameters of the
fits proved to be very sensitive to the number of harmonics and linear
combinations concerned, we used only those components which actually
appeared in the Fourier spectrum.

The modal content of the pulsation is quantified by the amplitude ratio
$A_1/A_0$: Fourier amplitude of the first overtone mode $A_1(=a_{01})$
divided by the amplitude of fundamental mode $A_0(=a_{10})$. We would
like to emphasize that the amplitude ratio defined above may differ
significantly from the ratio of the total amplitudes which is generally
used in the literature (e.g. \citealt{NC89}, \citealt{CCRRd}).

The estimation of errors of amplitude ratios yields the key to
justifying their possible changes. The systematic errors of the
amplitude ratios were calculated by a simulation: synthetic time series
were generated using the actual frequencies, amplitudes, phases of
reference light curves with observational times of data sets
from which mode
switching were published.  Fourier parameters were determined in the
same way as for the reference data in 1000 simulations/each data
set/each star. The initial epoch of simulated light curves ($T_0$) was
shifted at each iteration steps with 0.01 days and white noise was also
added with variance corresponding to the observed data. The mean
results of these tests are plotted in Fig.~\ref{sym}.
\begin{itemize}
\item
The errors of the amplitude ratios could be an order of magnitude
higher than their fitting errors.
\item
The same sampling rate yields strongly different accuracy of amplitude
ratio from star by star. Therefore, improving of time coverage of data
reduces the systematic error but not equally.
\item
The amplitude ratios were severely affected by the value of the initial
phase in those cases when only few nights of observations were involved
even in high precision CCD observations.
\end{itemize}
These results suggest that the formerly published mode switching events
were over-interpretations of badly sampled observations.

This conclusion are also supported by the theoretical calculations of
\citet{KB} who showed that the mode-switching time scale is about
hundred years in the most dramatic case.

\begin{table}
\caption[]{The colour dependence of the amplitude ratios of
multicolour CCD observations.
$(A_1/A_0)_B$, $(A_1/A_0)_V$, $(A_1/A_0)_I$ denote the
measured amplitude
ratios in Johnson\---Cousins {\it BVI\/} colours. Column Ref.~show
the sources of used data published by others.}
\centering
\begin{tabular}{lllll}
\hline
Star & $(A_1/A_0)_B$ & $(A_1/A_0)_V$ & $(A_1/A_0)_I$ & Ref.  \\
\hline
 V68  & $1.21\pm0.06$    & $1.21\pm0.05$  & $1.16\pm0.23$ & 1,3\\
 V68  & $1.25\pm0.06$    & $1.23\pm0.03$  & $1.14\pm0.05$ & 2,4,5\\
 V79  & $1.36\pm0.12$    & $1.37\pm0.12$  &               & 3\\
 V79  & $1.72\pm0.06$    & $1.87\pm0.05$  & $1.80\pm0.05$ & 2,4,5  \\
 V87  & $2.45\pm0.18$    & $2.52\pm0.25$  &               & 1,3 \\
 V87  & $2.62\pm0.15$    & $2.28\pm0.05$  & $2.25\pm0.08$ &  4,5  \\
 V99  & $1.88\pm0.07$    & $1.48\pm0.02$  &               &  3,4,5   \\
 V166   & $0.72\pm0.18$  & $0.78\pm0.09$  & $0.59\pm0.25$ & 3 \\
 V166   & $0.65\pm0.15$  & $0.60\pm0.05$  & $0.62\pm0.05$ & 5 \\
 V200   & $1.11\pm0.08$   & $1.01\pm0.07$ & $1.13\pm0.12$ & 5 \\
 V251   & $0.49\pm0.15$   & $0.81\pm0.10$   &             & 5 \\
 V252   & $1.14\pm0.05$   & $1.10\pm0.03$   &             & 4,5 \\
\hline
\end{tabular}

{\footnotesize \rm Ref.: 1. \citet{C98},
2. \citet{K98}, 3. \citet{CC01}, 4. \citet{hartman}, 5. this study.}
\label{ampcolor}
\end{table}

\begin{figure*}
\includegraphics[width=8.5cm]{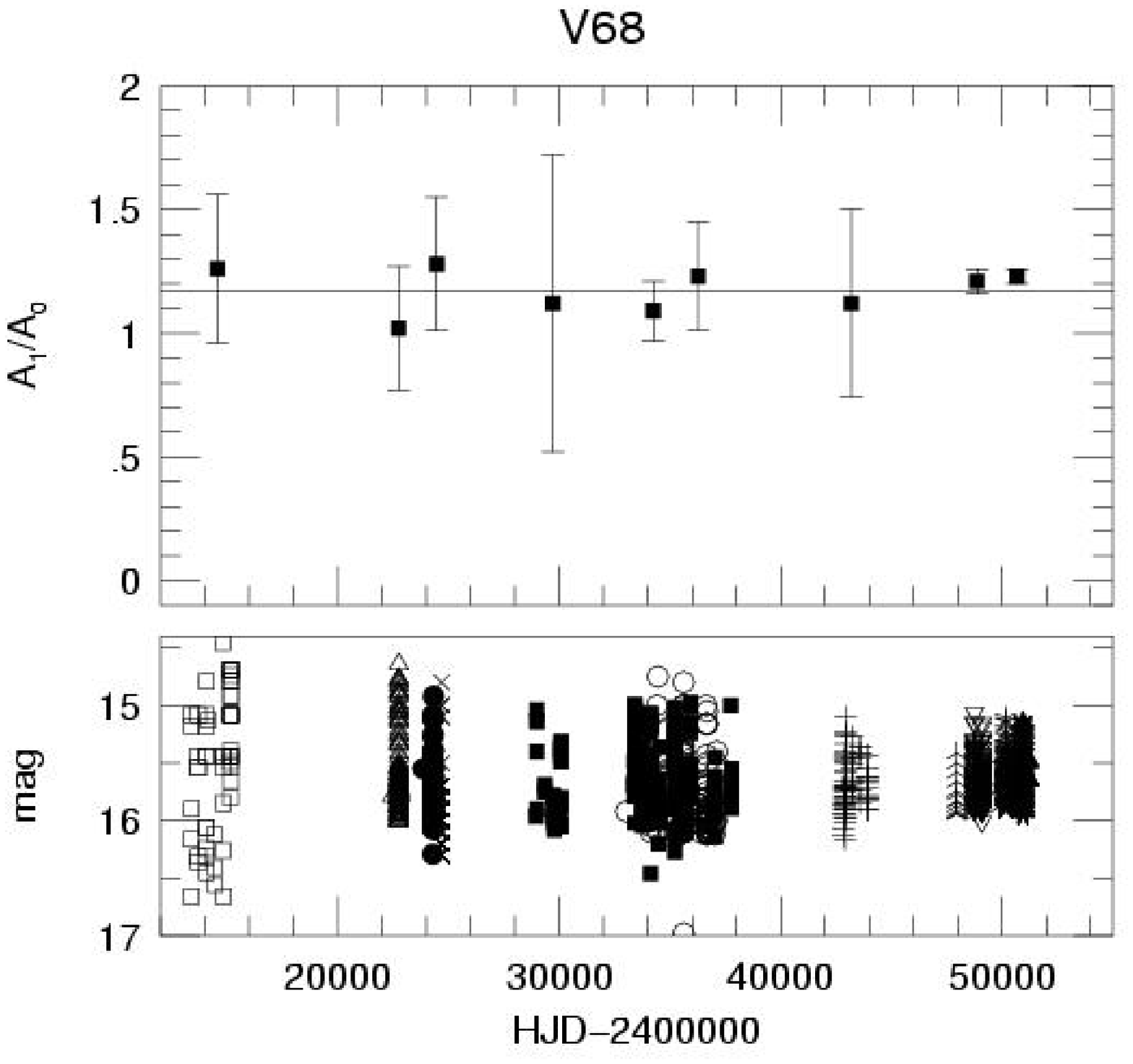}
\includegraphics[width=8.5cm]{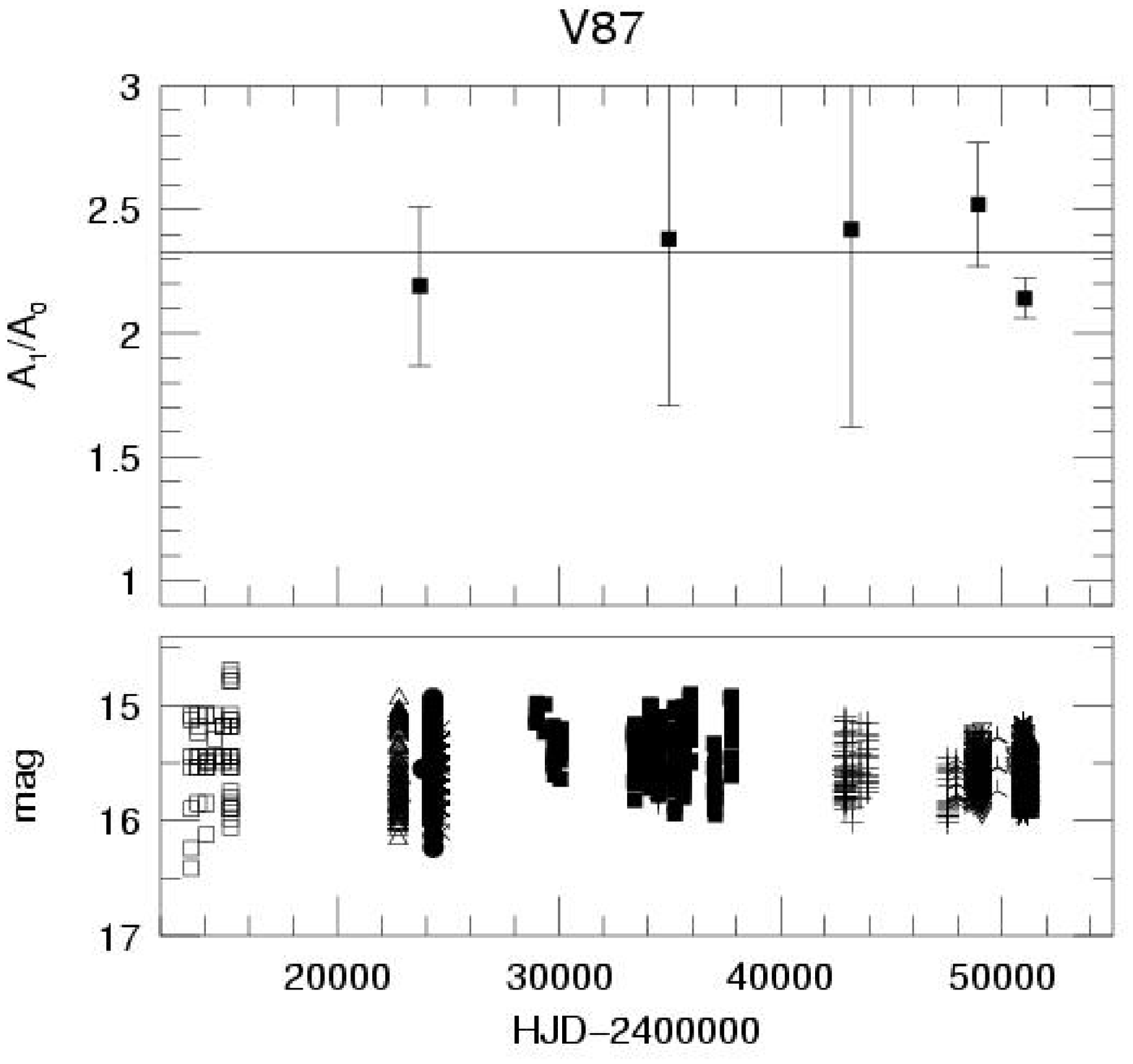}

\includegraphics[width=8.5cm]{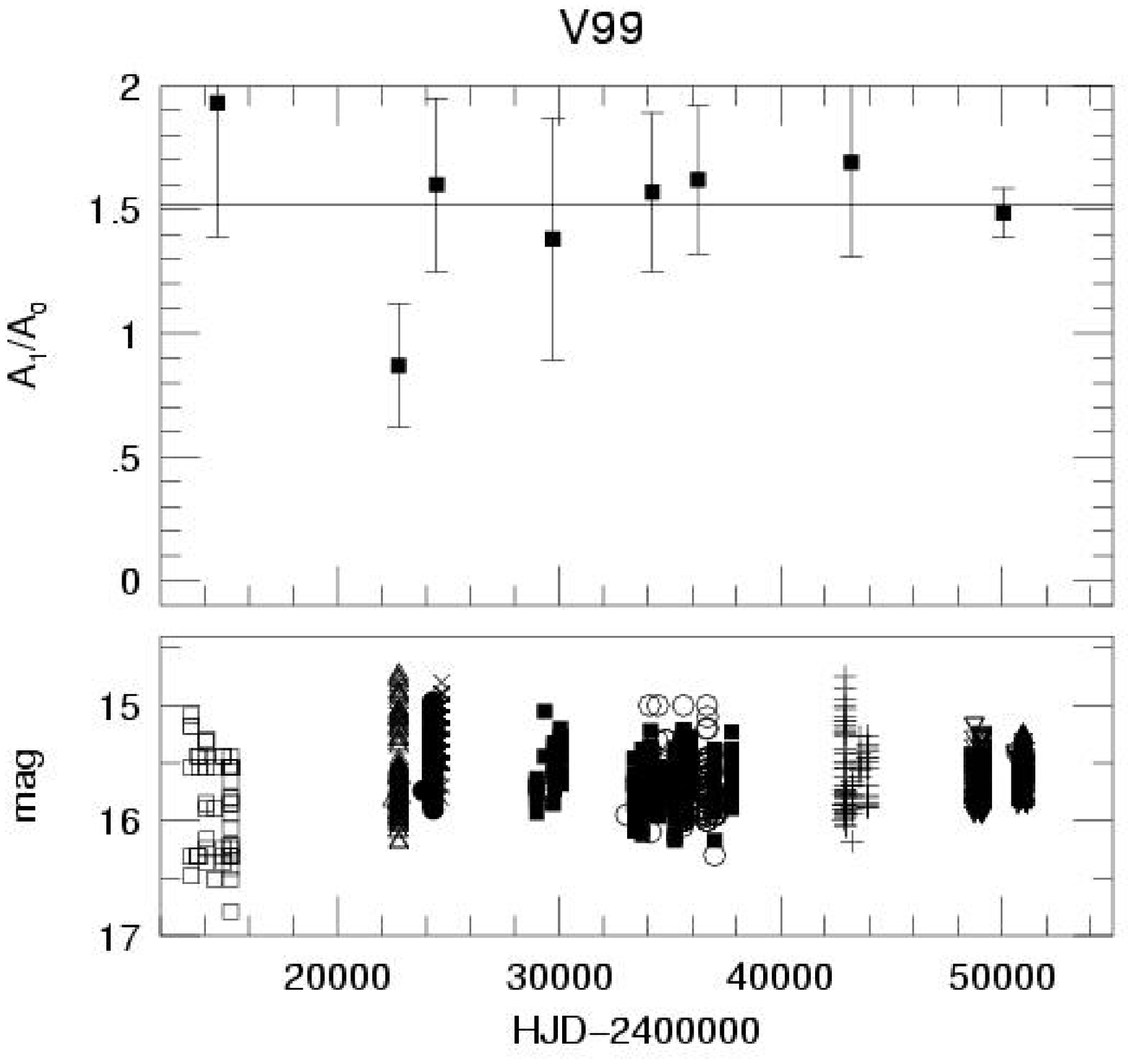}
\includegraphics[width=8.5cm]{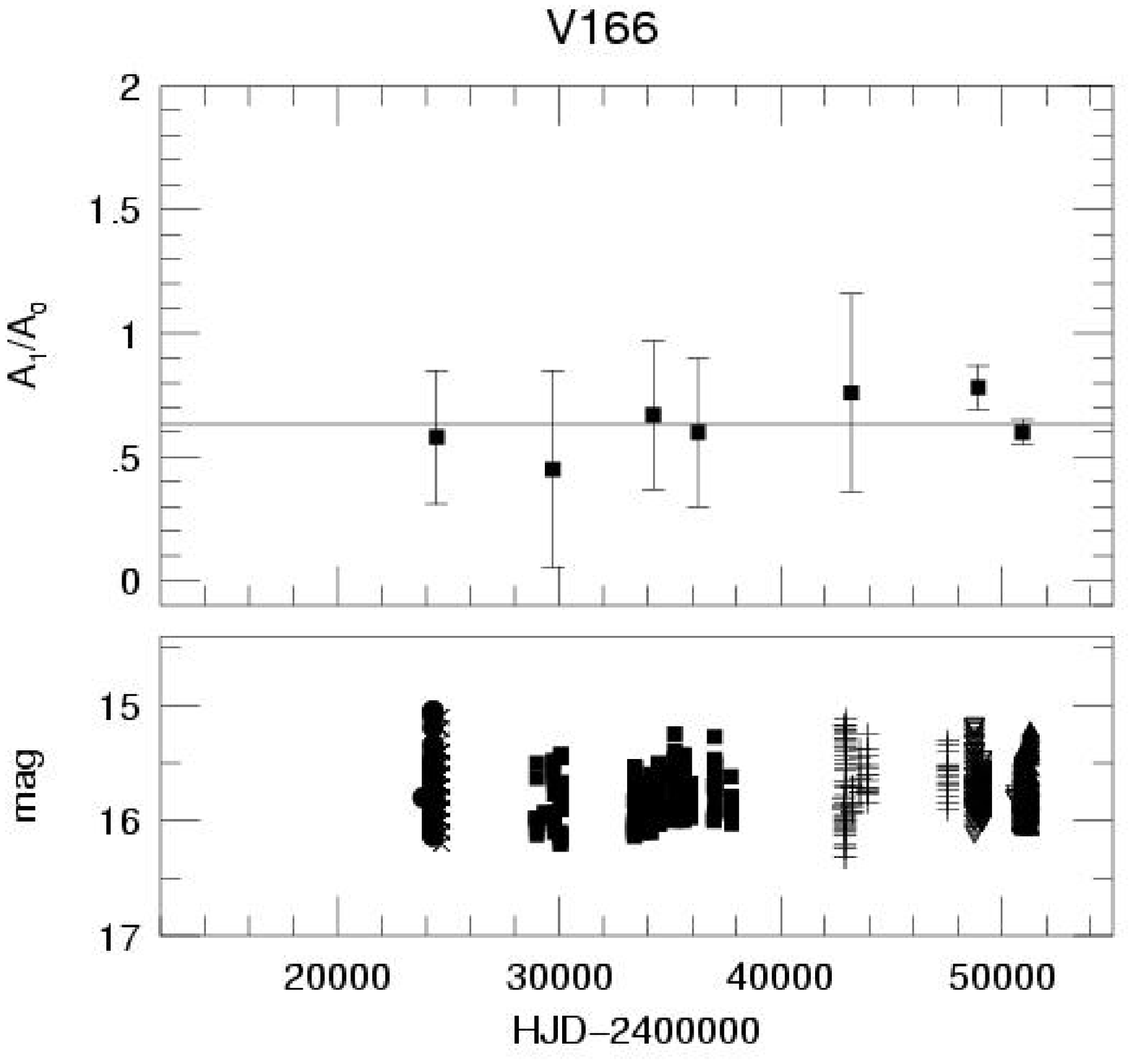}

\caption[]{Upper panels: amplitude ratios of the RRd variables in M3
determined for the different group of the data. The error-bars denote
the rms error of Fourier fits. Lower panels: the time distributions of
the photometric observations of each variables. Symbols: open squares:
\citet{Bailey}, open triangles: \citet{Larink}, dots: \citet{Mueller},
crosses: \citet{Greenstein}, filled squares: \citet{Bela65, Bela73},
open circles: \citet{Kuk61, Kuk70}, letters Y: \citet{RS}, pluses:
Konkoly Observatory (unpublished), trident stars: \citet{C98},
pentagrams: \citet{K98}, open triangles in upside down: \citet{CC01},
stars: \citet{hartman}, filled triangles: present work.}
\label{amplitude}
\end{figure*}

\subsubsection{Long time-scale stability of modes}

To investigate the mode switching phenomenon in longer time-scale all
published photometry (both photographic and CCD) was collected. Six RRd
stars (V13, V68, V79, V87, V99, V166) have sufficient amount of
historical data to study the possible long term changes of modal
content.  In the case of V13 and V79 no alternate periods have been
found from any parts of historical data while due to their central
position V200, V251 and V252 have not any old observations at all.  For
a practical review of published measurements of the remaining four
stars see the lower panels of Fig.~\ref{amplitude} where all different
data sets are plotted denoting by different symbols.

The data were divided into groups containing as many measurements as
possible in a moderate lengths of time interval where the periods seem
to be constants.  When different photometric observations belonged to
the same group, zero point corrections were applied to match the
photometries.

The effect of colour inhomogeneities on the amplitude ratios were
tested by determining them using the different colours of all the
available multicolour observations.  The results are shown in
Table~\ref{ampcolor}. It can be seen, that the colour dependence of the
amplitude ratio is generally smaller than the other errors entering in
the results.

Each data group were analyzed and fitted in the same manner as it was
described in Sec.~\ref{RRds}. The only difference is the number of used
Fourier harmonics in the case of photographic measurements (2-3).

In the top panels of the Figure~\ref{amplitude} it can be seen that all
studied RRd stars have a well defined amplitude ratio within the fitting
error. Neither mode switching nor severe amplitude ratio modification were
found.

\subsection{Notes on individual stars}\label{ind}

\noindent

{\bf V4} Two RRab stars are separated by less than $0\farcs5$
\citep{Gua} with periods of 0.585029 and 0.593069 days, respectively.

{\bf V5} The only star for which period of Blazhko cycle was estimated
previously.  The value of 194.551 days given by
\citet{Panov} seems to be not valid for CCD data. Unfortunately,
determining correct period was unsuccessful.

{\bf V10} As opposed to C98 we can confirm light curve variations
reported by
\citet{Bela65}. The star has a clearly visible Blazhko type amplitude
modulation.

{\bf V18} Although, by earlier catalogues this star was classified as a
Blazhko type RR Lyrae we could not detect such a behaviour on the basis
of all available data sets (K98, CC01, H05 and present work).

{\bf V22} The light curve variations are certain from both available
CCD photometries (CC01 and this work).

{\bf V23} Its brightness in band {\it I\/} is too high.

{\bf V29} The star is close to the V155. The period given by C98 is an
alias. CC01 interchanged V155 with V29. V29 is a short period RRab with
strong phase modulation.

{\bf V54} We agree with \citet{Bela65} that the star shows the Blazhko
effect clearly, albeit by newer sources (e.g. \citealt{Cl}, CC01) this
fact was neglected.

{\bf V44} CC04 reclassified it as an RRc variable star, but we agree with
the earlier work CC01 and the classification type as RRab with Blazhko
modulation.

{\bf V70} The star lies definitely above the horizontal branch and it
has an RRc like light curve with an unusually long period. It might be
an evolved RR Lyrae stars like V129, V170 and V261.  We cannot find a
common period with any of the C98, K98 nor CC01. This indicates strong
period changes.

{\bf V77} Unusually short period RRab type star.
As it was noted by CC01 its
colour index {\it B$-$V\/} placed them slightly to the RRc side of the
CMD.

{\bf V95} This variable star was reported by the ROTSE~I survey as a new
one by \citet{ROTSE} as ROTSE1 J134159.98+282257.0.

{\bf V104} Combining CCD data (K98, CC01 and ours) on this star its
Blazhko behaviour is obvious.

{\bf V114} Very few points. The star is lying far from the centre and
out of most of our frames.

{\bf V122} See comments at V229.

{\bf V129} Similar to V70.

{\bf V140} It has a light curve modulation both in phase and
amplitude as it was already mentioned by \citet{Bela65}.
Most probably this star is an RRc with Blazhko effect.

{\bf V146} C98, CC01 and CC04 have measured the V222 instead of this
star.

{\bf V149} The light curve variability is evident both from ours and
from CC01's data.

{\bf V155} CC01 have interchanged this star with V29.

{\bf V168} In addition to V140 this star is another example
for the Blazhko type RRc variables in the cluster.

{\bf V170} It is a variable star like V70. Unfortunately, it is merging
with V136.

{\bf V172} We have found the period to be much shorter ($\sim0.543$ d)
than it was given by The Catalogue (0.594 d) unless they accidentally
reversed the last two digits. A possible phase
modulation was detected as well.

{\bf V178} The pulsation frequency of the nearby V152 can be
recognized  clearly in the Fourier spectrum of V178.
Modulation of V178 is
probably caused by this neighbourhood.

{\bf V191} We reported here its evident Blazhko behaviour for the first
time.

{\bf V207} S02 published a period 0.4291 d. We agree with CC01 in a
much shorter one: 0.345307 d.

{\bf V211} The Blazhko effect noted in Table~\ref{table4} is clearly
seen in its phase diagram.

{\bf V218} shows Blazhko effect with large amplitude modulation.

{\bf V221} S02 have suggested its period as 0.6098 d. Folding our data
with the period 0.3787 d given by CC01 produced a slightly smaller
scatter.  The colour indices are those of type RRc stars.

{\bf V222} CC01 and CC04 have interchanged this star with V146.

{\bf V229} S02 obtained period 0.6877 d. We have found a much shorter
one: 0.49917 d.

{\bf V234} CC01 measured this star under the name of V164 as it was
corrected by CC04. S02 found a very different period: 0.5495 d.

{\bf V239} CC01 published a period as 0.333431 d. CC04 corrected it
to 0.66949 d and S02 found a similar one (0.6766 d). Nevertheless this
period is an alias! Preparing the phase diagram to this period two
maxima and minima are seen within one cycle. Period 0.503982 d from
our data yields correct phase diagrams for all data sets.

{\bf V242} Period of S02 0.6513 d is significantly different from the
common period of CC04 and ours.

{\bf V250} As it was shown in Section~\ref{new}, actually it is
two overlapping variable stars.

{\bf V259} was found by S02 as an RRab star with the period of about a
half a day. Our data fit is better with the RRc period of CC01 and CC04.

{\bf V261} is also an evolved star like V70, V129 and V170.

{\bf V266} was not included in any of former time series studies. The
period determined from our data suggest an RRc type star. The shape of
the folded light curve shows also a normal first overtone mode
pulsating RR Lyrae.

{\bf V270} It is two blended variable stars.
See Section~\ref{new} for details.

\section{Summary}

We have conducted the most extensive CCD {\it BVI\/} survey of variables
in the globular cluster M3
covering 238 stars.
Image subtraction photometric data transformed into the Johnson-Cousins
magnitude scale, light curves, periods and average magnitudes were
presented

\begin{itemize}
\item
{\it Varibility search\/}
In addition to our new variable stars  published in \citet{Acta},
we have discovered three further RR Lyrae stars.
We also found four non-variable stars
in the instability strip.  Period, brightness were
improved for most of the stars and several confused identifications were
clarified.

\item
{\it Color \--- magnitude diagrams\/}
At least three significant subgroups have been separated by
hierarchical clustering method in the CMDs of the RRab stars.  The
identified three clusters agree well with previous studies of
\citet{letter} and with CC05 when we contract the appropriate
clusters to 2 groups from our three ones or 3 from the four subgroups
separated by
\citet{letter}.

\item
{\it Period \--- brightness relations\/}
A clear correlation have been found between period of RRab stars and
their $\langle I \rangle$ magnitudes. The relation is in good agreement
with synthetic horizontal branch calculation published by \citet{Catelan}.

\item
{\it Modulated RR Lyrae stars\/}
We identified 66 modulated (Blazhko) RR Lyrae stars in our sample
where five
of them are new discoveries. For six stars the ambiguous Blazhko
classification have been verified. Two modulated RRc variable stars were
found in the cluster. By combining the available CCD data
modulation periods were estimated for 13 Blazhko star for the first time.

\item
{\it Double mode RR Lyraes\/}
Investigating the Fourier spectra of V252 its double-mode pulsation has
been confirmed.  Analyzing all available photometric data on RRd stars
of the cluster no undoubted modal content change, neither on shorter nor
longer time scales have been detected except for the case of V79. It
was shown, that to detect a mode-switching it is practically impossible
from available CCD photometric materials containing observing runs
from 1-3 seasons.
\end{itemize}

At the end of this paper we would like to call the attention to the
necessity of further long CCD time series observation of the variable
stars of M3. To answer the question of modal changes of the
double-mode variable stars or to find correct period for all modulated
stars and many similar problems need long-term, homogeneous and high
precision photometry.

\section*{Acknowledgments}

Parts of the data published here have been observed
by J. Jurcsik and G. Kov\'acs.
The authors wish to express their gratitude to them for providing
their data. JMB is grateful to J. Jurcsik,
S. Barcza and the referee for their comments and
suggestions and thank OTKA Grant No~T-43504 for partial financial support.
Work of G\'AB was partially supported by the NASA
Hubble Fellowship grant HST-HF-01170.01-A.

\bsp

\onecolumn

\begin{longtable}{@{}lllllllllll}
\caption{The basic parameters of the variables in M3.} \\
\hline 
\\[-7pt]
ID & Period & Type &
$\overline{B} $ & $\langle B \rangle$ &
$\overline{V} $ & $\langle V \rangle$ & 
$\overline{I} $ & $\langle I \rangle$ &
 ${\rm Ref}_{B, V, I}$ &  Comm. \\
\\[-7pt]
\hline
\\[-7pt]
\endfirsthead
\caption[]{(continued)}\\
\hline
\\[-7pt]
ID & Period & Type &
$\overline{B}$ & $\langle B \rangle$ &
$\overline{V}$ & $\langle V \rangle$ & 
$\overline{I}$ & $\langle I \rangle$ &
 ${\rm Ref}_{B, V, I}$ &  Comm. \\
\\[-7pt]
\hline
\\[-7pt]
\endhead
1	&	0.5205963	&	RRab	&	16.004	&	15.900	&	15.688	&	15.631	&	15.248	&	15.227	&	R	R	R	&	3\\ 
3	&	0.5581979	&	RRab	&	15.939	&	15.857	&	15.661	&	15.612	&	15.175	&	15.155	&	R	R	R	&	Bl,3\\ 
4n	&	0.585029	&	RRab	&	 	&	 	&	 	&	 	&	 	&	 	&	R	R	R	&	c\\ 
4s	&	0.593069	&	RRab	&	 	&	 	&	 	&	 	&	 	&	 	&	R	R	R	&	c\\ 
5	&	0.505703	&	RRab	&	15.986	&	15.974	&	15.725	&	15.714	&	15.243:	&	15.238:	&	A	A	S	&	Bl,c\\ 
6	&	0.5143327	&	RRab	&	16.083	&	16.003	&	15.755	&	15.703	&	15.294	&	15.276	&	R	R	R	&	1\\ 
7	&	0.4974248	&	RRab	&	15.980	&	15.889	&	15.694	&	15.638	&	15.248	&	15.230	&	R	R	R	&	Bl,2\\ 
8	&	0.636728	&	RRab	&	15.808	&	15.787	&	15.649	&	15.634	&	14.902	&	14.898	&	R:	R:	R:	&	3\\ 
9	&	0.5415553	&	RRab	&	16.031	&	15.957	&	15.689	&	15.645	&	15.233	&	15.217	&	A	A	S	&	2\\ 
10	&	0.5695465	&	RRab	&	16.057	&	16.004	&	15.684	&	15.647	&	15.162	&	15.152	&	S	S	S	&	Bl,2,c\\ 
11	&	0.5078915	&	RRab	&	15.993	&	15.873	&	15.706	&	15.636	&	15.188	&	15.166	&	S	S	S	&	3\\ 
12	&	0.3179347	&	RRc	&	15.840	&	15.814	&	15.621	&	15.606	&	15.281	&	15.275	&	R	R	R	&	 \\ 
13	&	0.4795043	&	RRd	&	15.931	&	15.917	&	15.691	&	15.684	&	15.243	&	15.239	&	R	R	R	&	$P_1=0.351579$\\ 
14	&	0.6359002	&	RRab	&	15.914	&	15.867	&	15.581	&	15.551	&	15.059	&	15.050	&	R	R	R	&	Bl,4\\ 
15	&	0.5300874	&	RRab	&	15.919	&	15.833	&	15.667	&	15.616	&	15.159	&	15.144	&	A	A	A	&	3\\ 
16	&	0.5114943	&	RRab	&	16.110	&	16.020	&	15.759	&	15.701	&	15.307	&	15.288	&	A	A	S	&	1\\ 
17	&	0.5761594	&	RRab	&	16.060	&	16.020	&	15.694	&	15.680	&	15.136	&	15.127	&	A	A	A	&	Bl\\ 
18	&	0.5164527	&	RRab	&	16.179:	&	16.068:	&	15.751	&	15.698	&	15.251:	&	15.231:	&	A	A	A	&	c,1\\ 
19	&	0.6319774	&	RRab	&	16.109	&	16.093	&	15.689	&	15.681	&	15.148	&	15.145	&	A	A	S	&	2\\ 
20	&	0.4912636	&	RRab	&	16.075	&	15.988	&	15.707	&	15.656	&	15.233	&	15.215	&	S	A	S	&	Bl\\ 
21	&	0.515758	&	RRab	&	16.103	&	16.012	&	15.742	&	15.690	&	 	&	 	&	S	S	S	&	a,2\\ 
22	&	0.4814205	&	RRab	&	16.085:	&	15.989:	&	15.762	&	15.698	&	15.292	&	15.269	&	S	A	S	&	Bl,c\\ 
23	&	0.5953834	&	RRab	&	15.934	&	15.884	&	15.679	&	15.639	&	14.683	&	14.677	&	S:	S:	S:	&	Bl,c\\ 
24	&	0.663380	&	RRab	&	15.961	&	15.914	&	15.558	&	15.533	&	15.068	&	15.060	&	A	S	S	&	Bl,4\\ 
25	&	0.4800623	&	RRab	&	16.002	&	15.886	&	15.747	&	15.681	&	15.294	&	15.271	&	R	R	R	&	2\\ 
26	&	0.5977427	&	RRab	&	15.962	&	15.891	&	15.639	&	15.596	&	15.002	&	14.991	&	A	S	S	&	3\\ 
27	&	0.5790682	&	RRab	&	16.031	&	15.972	&	15.679	&	15.649	&	15.242	&	15.230	&	R	R	R	&	2\\ 
28	&	0.470617	&	RRab	&	15.982	&	15.911	&	15.709	&	15.655	&	15.281	&	15.268	&	R	R	R	&	Bl\\ 
29	&	0.4717879	&	RRab	&	15.918	&	15.874	&	15.821	&	15.769	&	15.271	&	15.265	&	R	R	R	&	Bl,m,c\\ 
30	&	0.5120888	&	RRab	&	15.993	&	15.895	&	15.726	&	15.676	&	15.265	&	15.248	&	R	R	R	&	m,2\\ 
31	&	0.5807219	&	RRab	&	15.946	&	15.829	&	15.612	&	15.556	&	15.160	&	15.140	&	R	R	R	&	4\\ 
32	&	0.4953506	&	RRab	&	16.043	&	15.924	&	15.736	&	15.675	&	15.265	&	15.244	&	R	R	R	&	2\\ 
33	&	0.5252299	&	RRab	&	15.944	&	15.877	&	15.705	&	15.670	&	15.200	&	15.191	&	R	R	R	&	Bl\\ 
34	&	0.5591031	&	RRab	&	16.089	&	16.067	&	15.787	&	15.768	&	15.179	&	15.177	&	A	A	S	&	Bl\\ 
35	&	0.5305483	&	RRab	&	15.998	&	15.878	&	15.695	&	15.620	&	15.132:	&	15.112:	&	A	A	S	&	Bl\\ 
36	&	0.5455929	&	RRab	&	15.944	&	15.854	&	15.660	&	15.601	&	15.235	&	15.214	&	S	A	S	&	3\\ 
37	&	0.3266378	&	RRc	&	15.928	&	15.903	&	15.693	&	15.679	&	15.307	&	15.302	&	S	A	S	&	 \\ 
38	&	0.5580110	&	RRab	&	16.108	&	16.086	&	15.765	&	15.701	&	15.191	&	15.171	&	A	A	S	&	Bl\\ 
39	&	0.5870758	&	RRab	&	16.093	&	16.021	&	15.701	&	15.668	&	15.220	&	15.206	&	A	A	S	&	Bl\\ 
40	&	0.5515394	&	RRab	&	16.110	&	16.038	&	15.711	&	15.672	&	15.231	&	15.218	&	A	A	S	&	2\\ 
41	&	0.4858777	&	RRab	&	16.115	&	16.001	&	15.712	&	15.659	&	15.311	&	15.292	&	R	R	R	&	Bl\\ 
42	&	0.5900938	&	RRab	&	15.854	&	15.724	&	15.563	&	15.497	&	15.103	&	15.079	&	R	R	R	&	4\\ 
43	&	0.5405129	&	RRab	&	16.089	&	16.006	&	15.719	&	15.671	&	15.244	&	15.228	&	R	R	R	&	Bl,1\\ 
44	&	0.5063801	&	RRab	&	15.972	&	15.912	&	15.659	&	15.615	&	15.171	&	15.160	&	A	A	S	&	Bl\\ 
45	&	0.5369002	&	RRab	&	16.118	&	16.056	&	15.706	&	15.675	&	15.259	&	15.247	&	S	A	S	&	Bl\\ 
46	&	0.6133832	&	RRab	&	16.050	&	16.028	&	15.716	&	15.703	&	15.158	&	15.153	&	R	R	R	&	1\\ 
47	&	0.5409128	&	RRab	&	15.995	&	15.941	&	15.673	&	15.648	&	15.200	&	15.190	&	R	R	R	&	Bl\\ 
48	&	0.6278299	&	RRab	&	15.952	&	15.922	&	15.631	&	15.615	&	 	&	 	&	S	S	S	&	3\\ 
49	&	0.5482048	&	RRab	&	 	&	 	&	15.696	&	15.645	&	 	&	 	&	S	A	S	&	Bl\\ 
50	&	0.5128392	&	RRab	&	16.038	&	15.991	&	15.674	&	15.645	&	15.173	&	15.165	&	S	A	S	&	Bl\\ 
51	&	0.5839702	&	RRab	&	16.093	&	16.041	&	15.688	&	15.658	&	15.194:	&	15.187:	&	S	S	A	&	2\\ 
52	&	0.5162299	&	RRab	&	 	&	 	&	15.773	&	15.756	&	 	&	 	&	R	R	R	&	Bl\\ 
53	&	0.5048789	&	RRab	&	16.052	&	15.939	&	15.731	&	15.669	&	15.301	&	15.281	&	R	R	R	&	2\\ 
54	&	0.5062570	&	RRab	&	16.102	&	16.044	&	15.753	&	15.714	&	15.356	&	15.344	&	R	R	R	&	Bl,c\\ 
55	&	0.5298207	&	RRab	&	16.063	&	15.976	&	15.724	&	15.673	&	 	&	 	&	A	A	S	&	2\\ 
56	&	0.3296000	&	RRc	&	15.885	&	15.866	&	15.658	&	15.642	&	15.257	&	15.253	&	A	A	S	&	 \\ 
57	&	0.5121885	&	RRab	&	16.078	&	16.001	&	15.748	&	15.699	&	15.378	&	15.357	&	S	S	S	&	2\\ 
58	&	0.5170549	&	RRab	&	15.936	&	15.825	&	15.670	&	15.600	&	15.218	&	15.196	&	R	R	R	&	3\\ 
59	&	0.5888236	&	RRab	&	16.130	&	16.071	&	15.677	&	15.649	&	15.179	&	15.170	&	A	A	S	&	Bl,2\\ 
60	&	0.7077280	&	RRab	&	15.970	&	15.936	&	15.525	&	15.506	&	 	&	 	&	S	A	S	&	4\\ 
61	&	0.5209053	&	RRab	&	16.062:	&	15.991:	&	15.733	&	15.690	&	15.220:	&	15.207:	&	S	A	S	&	Bl\\ 
62	&	0.6524099	&	RRab	&	16.055	&	16.027	&	15.657	&	15.643	&	15.091	&	15.085	&	A	S	S	&	Bl,3\\ 
63	&	0.5704004	&	RRab	&	16.081	&	16.052	&	15.710	&	15.670	&	15.170	&	15.163	&	S	A	S	&	Bl\\ 
64	&	0.6054588	&	RRab	&	16.080	&	16.045	&	15.693	&	15.672	&	15.162:	&	15.156:	&	A	S	S	&	2\\ 
65	&	0.6683474	&	RRab	&	15.940	&	15.875	&	15.531	&	15.497	&	15.004	&	14.991	&	A	S	A	&	4\\ 
66	&	0.6201807	&	RRab	&	16.026	&	15.991	&	15.659	&	15.641	&	15.150	&	15.142	&	R	R	R	&	Bl\\ 
67	&	0.5683324	&	RRab	&	16.058	&	15.976	&	15.683	&	15.635	&	15.218	&	15.200	&	R	R	R	&	Bl\\ 
68	&	0.3559911	&	RRd	&	15.929	&	15.916	&	15.636	&	15.632	&	15.267	&	15.265	&	R	R	R	&	$P_0=0.478585$\\ 
69	&	0.5666159	&	RRab	&	16.101	&	16.037	&	15.704	&	15.672	&	15.219	&	15.208	&	R	R	R	&	2\\ 
70	&	0.4863518	&	RRc(?)	&	15.684	&	15.670	&	15.392	&	15.386	&	14.956	&	14.953	&	R	R	R	&	c\\ 
71	&	0.5490530	&	RRab	&	16.048	&	16.008	&	15.825	&	15.796	&	15.033	&	15.024	&	A:	S:	A:	&	Bl?,m\\ 
72	&	0.4560780	&	RRab	&	16.076	&	15.953	&	15.743	&	15.670	&	15.349	&	15.320	&	A	A	S	&	2\\ 
73	&	0.673611	&	RRab	&	16.054	&	16.047	&	15.629	&	15.626	&	15.093	&	15.092	&	S	A	S	&	3\\ 
74	&	0.4921525	&	RRab	&	16.129	&	15.994	&	15.759	&	15.692	&	15.325	&	15.304	&	R	R	R	&	2\\ 
75	&	0.3140767	&	RRc	&	15.914	&	15.890	&	15.640	&	15.626	&	15.325	&	15.319	&	R	R	R	&	 \\ 
76	&	0.5017627	&	RRab	&	16.076	&	15.967	&	15.778	&	15.720	&	15.317:	&	15.294:	&	R	R	R	&	1\\ 
77	&	0.4593499	&	RRab	&	15.980	&	15.870	&	15.816	&	15.743	&	15.333	&	15.311	&	R	R	R	&	Bl,c,1\\ 
78	&	0.61192	&	RRab	&	15.969	&	15.944	&	15.613	&	15.602	&	15.117	&	15.110	&	R	R	R	&	Bl\\ 
79	&	0.3580727	&	RRd	&	15.999	&	15.988	&	15.749	&	15.742	&	15.268	&	15.266	&	A	A	S	&	$P_0=0.478814$\\ 
80	&	0.5383788	&	RRab	&	16.027	&	15.955	&	15.679	&	15.632	&	15.218	&	15.205	&	A	A	S	&	Bl\\ 
81	&	0.5291220	&	RRab	&	16.078	&	15.986	&	15.730	&	15.679	&	15.277:	&	15.255:	&	A	A	S	&	2\\ 
83	&	0.5012631	&	RRab	&	16.057:	&	15.960:	&	15.733	&	15.674	&	15.248:	&	15.228:	&	A	A	A	&	2\\ 
84	&	0.5957289	&	RRab	&	16.082	&	16.042	&	15.678	&	15.658	&	15.166	&	15.158	&	R	R	R	&	2\\ 
85	&	0.3558191	&	RRc	&	15.780	&	15.754	&	15.591	&	15.574	&	15.180	&	15.176	&	A	S	S	&	 \\ 
86	&	0.2926596	&	RRc	&	15.937	&	15.909	&	15.691	&	15.673	&	15.381	&	15.375	&	S	A	S	&	 \\ 
87	&	0.3574779	&	RRd	&	15.870	&	15.856	&	15.580	&	15.571	&	15.190	&	15.187	&	R	R	R	&	$P_0=0.480229$\\ 
88	&	0.2987499	&	RRc	&	15.900	&	15.868	&	15.706	&	15.687	&	15.183	&	15.179	&	R	R	R	&	 \\ 
89	&	0.5484803	&	RRab	&	16.040	&	15.955	&	15.689	&	15.647	&	15.190	&	15.176	&	R	R	R	&	2\\ 
90	&	0.5170308	&	RRab	&	16.111:	&	16.008:	&	15.727	&	15.673	&	15.207:	&	15.190:	&	S	A	S	&	2\\ 
92	&	0.5035471	&	RRab	&	16.041	&	15.949	&	15.723	&	15.672	&	15.222	&	15.201	&	A	A	S	&	a,2\\ 
93	&	0.6022960	&	RRab	&	16.060	&	16.021	&	15.662	&	15.640	&	15.151	&	15.143	&	A	A	S	&	2\\ 
94	&	0.5236940	&	RRab	&	16.082	&	15.993	&	15.725	&	15.671	&	15.252	&	15.235	&	A	A	S	&	2\\ 
95	&	103.5	&	long	&	 	&	 	&	 	&	 	&	 	&	 	&	A	A	A	&	c\\ 
96	&	0.4994150	&	RRab	&	16.023	&	15.918	&	15.719	&	15.652	&	 	&	 	&	A	S	S	&	a,2\\ 
97	&	0.3349326	&	RRc	&	15.947	&	15.933	&	15.712	&	15.703	&	15.359	&	15.355	&	S	S	S	&	 \\ 
99	&	0.3609319	&	RRd	&	15.832:	&	15.824:	&	15.592	&	15.590	&	15.098:	&	15.096:	&	S	A	S	&	$P_0=0.482006$\\ 
100	&	0.6188126	&	RRab	&	16.074	&	16.045	&	15.721	&	15.705	&	15.171	&	15.165	&	R	R	R	&	1\\ 
101	&	0.6438879	&	RRab	&	16.117	&	16.093	&	15.703	&	15.692	&	15.169	&	15.166	&	R	R	R	&	Bl,2\\ 
104	&	0.5699259	&	RRab	&	15.883	&	15.773	&	15.592	&	15.536	&	15.164	&	15.145	&	R	R	R	&	Bl,c,4\\ 
105	&	0.2877440	&	RRc	&	15.802	&	15.791	&	15.586	&	15.580	&	15.308	&	15.306	&	R	R	R	&	 \\ 
106	&	0.5471316	&	RRab	&	16.112	&	16.039	&	15.712	&	15.675	&	15.237	&	15.223	&	R	R	R	&	Bl\\ 
107	&	0.3090378	&	RRc	&	15.917	&	15.890	&	15.690	&	15.674	&	15.299	&	15.293	&	A	A	S	&	 \\ 
108	&	0.519610	&	RRab	&	16.093	&	16.000	&	15.762	&	15.708	&	15.260	&	15.241	&	S	S	S	&	a,1\\ 
109	&	0.5339205	&	RRab	&	16.030	&	15.945	&	15.746	&	15.695	&	15.227	&	15.211	&	R	R	R	&	1\\ 
110	&	0.5354602	&	RRab	&	16.035	&	15.959	&	15.747	&	15.698	&	15.222	&	15.206	&	R	R	R	&	Bl\\ 
111	&	0.5101919	&	RRab	&	16.007	&	15.938	&	15.754	&	15.694	&	15.314	&	15.290	&	R	R	R	&	Bl\\ 
114	&	0.597723	&	RRab	&	 	&	 	&	 	&	 	&	 	&	 	&	A	A	A	&	c\\ 
116	&	0.514811	&	RRab	&	16.116	&	16.026	&	15.762	&	15.706	&	 	&	 	&	A	A	S	&	a,1\\ 
117	&	0.6005350	&	RRab	&	15.986	&	15.913	&	15.642	&	15.592	&	15.059	&	15.042	&	A	A	A	&	Bl\\ 
118	&	0.4993905	&	RRab	&	16.092	&	15.987	&	15.730	&	15.672	&	 	&	 	&	A	A	S	&	a,2\\ 
119	&	0.517690	&	RRab	&	16.062:	&	15.966:	&	15.729	&	15.670	&	15.201:	&	15.182:	&	A	A	S	&	2\\ 
120	&	0.640140	&	RRab	&	16.023	&	16.007	&	15.637	&	15.628	&	15.093	&	15.090	&	A	A	S	&	3\\ 
121	&	0.5352076	&	RRab	&	16.050	&	16.016	&	15.736	&	15.704	&	15.292	&	15.286	&	R	R	R	&	Bl\\ 
122	&	0.5159519	&	RRab	&	 	&	 	&	 	&	 	&	 	&	 	&	R	R	R	&	m,c\\ 
124	&	0.752439	&	RRab	&	16.012	&	16.000	&	15.550	&	15.544	&	14.944	&	14.942	&	S	S	S	&	4\\ 
125	&	0.3498227	&	RRc	&	15.978	&	15.959	&	15.740	&	15.730	&	15.224	&	15.220	&	S	S	S	&	 \\ 
126	&	0.3484062	&	RRc	&	15.903	&	15.887	&	15.631	&	15.622	&	15.275	&	15.271	&	R	R	R	&	 \\ 
128	&	0.2920410	&	RRc	&	15.864	&	15.840	&	15.653	&	15.637	&	15.350	&	15.344	&	R	R	R	&	 \\ 
129	&	0.4060852	&	RRc(?)	&	15.778	&	15.760	&	15.500	&	15.492	&	15.073	&	15.070	&	R	R	R	&	c\\ 
130	&	0.5660594	&	RRab	&	15.972	&	15.958	&	15.679	&	15.671	&	15.160	&	15.156	&	R	R	R	&	Bl\\ 
131	&	0.2976890	&	RRc	&	15.904	&	15.878	&	15.695	&	15.679	&	15.354	&	15.349	&	R	R	R	&	 \\ 
132	&	0.3398560	&	RRc	&	15.912	&	15.895	&	15.632	&	15.622	&	15.251	&	15.249	&	R	R	R	&	 \\ 
133	&	0.5507236	&	RRab	&	16.048	&	15.977	&	15.734	&	15.701	&	15.275	&	15.261	&	R	R	R	&	Bl,2\\ 
134	&	0.6180597	&	RRab	&	15.995	&	15.971	&	15.680	&	15.663	&	15.145	&	15.139	&	R	R	R	&	2\\ 
135	&	0.5683920	&	RRab	&	16.006	&	15.965	&	15.743	&	15.706	&	15.239	&	15.228	&	R	R	R	&	m\\ 
136	&	0.617179	&	RRab	&	15.986	&	15.957	&	15.632	&	15.616	&	15.167	&	15.161	&	R	R	R	&	m,3\\ 
137	&	0.5751482	&	RRab	&	16.001	&	15.950	&	15.651	&	15.619	&	15.208	&	15.196	&	R	R	R	&	3\\ 
138	&	53.28	&	long	&	 	&	 	&	 	&	 	&	 	&	 	&	A	A	A	&	 \\ 
139	&	0.560000	&	RRab	&	15.961	&	15.841	&	15.648	&	15.587	&	15.184	&	15.160	&	R	R	R	&	4\\ 
140	&	0.3331388	&	RRc	&	15.740	&	15.715	&	15.542	&	15.530	&	15.228	&	15.225	&	R	R	R	&	Bl,c\\ 
142	&	0.5686274	&	RRab	&	16.100	&	16.029	&	15.713	&	15.679	&	15.279	&	15.266	&	R	R	R	&	2\\ 
143	&	0.5965350	&	RRab	&	15.749	&	15.662	&	15.437	&	15.396	&	15.010	&	14.993	&	R	R	R	&	Bl\\ 
144	&	0.5967843	&	RRab	&	16.042	&	16.000	&	15.698	&	15.675	&	15.144	&	15.136	&	R	R	R	&	2\\ 
145	&	0.5144880	&	RRab	&	15.825	&	15.752	&	15.539	&	15.499	&	14.972	&	14.960	&	R	R	R	&	m\\ 
146	&	0.5021930	&	RRab	&	16.006	&	15.905	&	15.734	&	15.674	&	15.266	&	15.246	&	R	R	R	&	c\\ 
147	&	0.3464809	&	RRc	&	15.964	&	15.943	&	15.733	&	15.722	&	15.304	&	15.300	&	R	R	R	&	 \\ 
148	&	0.4672693	&	RRab	&	16.173	&	16.026	&	15.823	&	15.752	&	15.366	&	15.343	&	R	R	R	&	m\\ 
149	&	0.5481796	&	RRab	&	16.083	&	15.992	&	15.679	&	15.641	&	15.277	&	15.250	&	R	R	R	&	Bl,c\\ 
150	&	0.5239248	&	RRab	&	16.066	&	15.978	&	15.737	&	15.687	&	15.256	&	15.238	&	R	R	R	&	Bl\\ 
151	&	0.5168247	&	RRab	&	15.991:	&	15.928:	&	15.647	&	15.621	&	15.273	&	15.260	&	R	R	R	&	Bl\\ 
152	&	0.3261308	&	RRc	&	15.823	&	15.805	&	15.551	&	15.540	&	15.294	&	15.289	&	R	R	R	&	 \\ 
154	&	15.2842	&	WVir	&	 	&	 	&	 	&	 	&	 	&	 	&	R	R	R	&	 \\ 
155	&	0.338050	&	RRc	&	15.815	&	15.805	&	15.667	&	15.660	&	15.257	&	15.254	&	R	R	R	&	m,c\\ 
156	&	0.531986	&	RRab	&	16.015	&	15.919	&	15.726	&	15.681	&	15.302	&	15.282	&	R:	R	R	&	2\\ 
157	&	0.542851	&	RRab	&	16.026	&	15.985	&	15.750	&	15.728	&	15.295	&	15.288	&	R	R	R	&	Bl\\ 
159	&	0.533890	&	RRab	&	15.733	&	15.678	&	15.407	&	15.380	&	14.914	&	14.903	&	R	R	R	&	m\\ 
160	&	0.657330	&	RRab	&	15.879	&	15.82	&	15.559	&	15.538	&	15.096	&	15.086	&	R	R	R	&	Bl,4\\ 
161	&	0.526495	&	RRab	&	16.076	&	15.966	&	15.754	&	15.693	&	15.245	&	15.221	&	R	R	R	&	Bl\\ 
165	&	0.4836315	&	RRab	&	15.920	&	15.802	&	15.647	&	15.585	&	15.244	&	15.222	&	R	R	R	&	 \\ 
166	&	0.4849529	&	RRd	&	15.986	&	15.969	&	15.809	&	15.797	&	15.275	&	15.271	&	R	R	R	&	$P_1=0.360147$\\ 
167	&	0.6439710	&	RRab	&	16.074	&	16.064	&	15.684	&	15.677	&	15.133	&	15.131	&	R	R	R	&	1\\ 
168	&	0.2759408	&	RRc	&	15.786	&	15.772	&	15.628	&	15.617	&	15.278	&	15.274	&	R	R	R	&	Bl,c\\ 
170	&	0.4323323	&	RRc(?)	&	15.613	&	15.590	&	15.416	&	15.406	&	15.044	&	15.039	&	R	R	R	&	m,c\\ 
171	&	0.303300	&	RRc	&	15.842	&	15.811	&	15.631	&	15.613	&	15.432	&	15.427	&	R	R	R	&	m\\ 
172	&	0.5422899	&	RRab	&	16.074	&	15.995	&	15.788	&	15.741	&	15.291	&	15.276	&	R	R	R	&	c,1\\ 
173	&	0.6070027	&	RRab	&	15.974	&	15.901	&	15.659	&	15.622	&	15.135	&	15.123	&	R	R	R	&	3\\ 
174	&	0.5913187	&	RRab	&	16.019	&	15.988	&	15.738	&	15.722	&	15.370	&	15.361	&	R	R	R	&	 \\ 
175	&	0.569697	&	RRab	&	16.092	&	16.026	&	15.699	&	15.665	&	15.183	&	15.169	&	R	R	R	&	2\\ 
176	&	0.5396109	&	RRab	&	16.197	&	16.128	&	15.733	&	15.707	&	15.235	&	15.224	&	R	R	R	&	Bl\\ 
177	&	0.3483497	&	RRc	&	15.699	&	15.671	&	15.534	&	15.517	&	15.270	&	15.264	&	R	R	R	&	 \\ 
178	&	0.2669625	&	RRc	&	15.905	&	15.891	&	15.705	&	15.699	&	15.480	&	15.478	&	R	R	R	&	m,c\\ 
180	&	0.6091004	&	RRab	&	16.050	&	16.017	&	15.702	&	15.679	&	15.347	&	15.336	&	R	R	R	&	Bl,2\\ 
181	&	0.6638463	&	RRab	&	15.893	&	15.870	&	15.540	&	15.528	&	 	&	 	&	R	R	 	&	 \\ 
184	&	0.5312073	&	RRab	&	15.971	&	15.948	&	15.690	&	15.660	&	 	&	 	&	R	R	 	&	m\\ 
186	&	0.6634230	&	RRab	&	15.944	&	15.931	&	15.619	&	15.613	&	15.100	&	15.095	&	R	R	R	&	3\\ 
187	&	0.586257	&	RRab	&	16.164	&	16.094	&	15.731	&	15.709	&	15.280	&	15.269	&	R	R	R	&	Bl?,1\\ 
188	&	0.26652823	&	RRc	&	15.958	&	15.944	&	15.789	&	15.782	&	15.530	&	15.527	&	R	R	R	&	 \\ 
189	&	0.61294	&	RRab	&	16.161	&	16.116	&	15.706	&	15.687	&	15.184	&	15.175	&	R	R	R	&	m,2\\ 
190	&	0.5227970	&	RRab	&	16.014	&	15.929	&	15.737	&	15.683	&	15.234	&	15.222	&	R	R	R	&	2\\ 
191	&	0.519206	&	RRab	&	15.996:	&	15.926:	&	15.736	&	15.702	&	15.333	&	15.321	&	R	R	R	&	Bl,c\\ 
193	&	0.747860	&	RRab	&	15.853	&	15.772	&	15.502	&	15.466	&	14.984	&	14.970	&	R:	R:	R:	&	m,4\\ 
194	&	0.4892000	&	RRab	&	15.910:	&	15.772:	&	15.679	&	15.628	&	15.270	&	15.255	&	R	R	R	&	 \\ 
195	&	0.64408	&	RRab	&	15.971	&	15.953	&	15.612	&	15.606	&	15.172:	&	15.169:	&	R	R	R	&	3\\ 
197	&	0.4998971	&	RRab	&	16.158	&	16.016	&	15.788	&	15.719	&	 	&	 	&	R	R	R	&	a,1\\ 
200	&	0.3610002	&	RRd	&	15.951	&	15.934	&	15.568	&	15.561	&	15.037	&	15.035	&	R	R	R	&	$P_0=0.485298$\\ 
201	&	0.54054	&	RRab	&	16.083	&	15.964	&	15.706	&	15.670	&	15.485	&	15.454	&	R	R	R	&	 \\ 
202	&	0.773571	&	RRab	&	15.961	&	15.959	&	15.559	&	15.558	&	14.977:	&	14.977:	&	A	A	S	&	4\\ 
203	&	0.2897940	&	RRc	&	15.858	&	15.855	&	15.593	&	15.592	&	15.253	&	15.252	&	S	A	S	&	 \\ 
207	&	0.345307	&	RRc	&	15.838	&	15.819	&	15.628	&	15.617	&	15.319	&	15.315	&	R	R	R	&	c\\ 
208	&	0.3383847	&	RRc	&	15.872	&	15.853	&	15.709	&	15.698	&	15.327	&	15.323	&	R	R	R	&	 \\ 
209	&	0.3482784	&	RRc	&	15.801	&	15.780	&	15.621	&	15.607	&	15.251	&	15.246	&	R	R	R	&	 \\ 
210	&	0.352947	&	RRc	&	15.837	&	15.820	&	15.683	&	15.672	&	15.230	&	15.223	&	R	R	R	&	 \\ 
211	&	0.558205	&	RRab	&	15.988	&	15.948	&	15.781	&	15.760	&	15.236	&	15.227	&	R:	R:	R:	&	Bl,m,c\\ 
212	&	0.5421882	&	RRab	&	15.985	&	15.919	&	15.659	&	15.622	&	15.429:	&	15.411:	&	R	R	R	&	Bl\\ 
213	&	0.299955	&	RRc	&	15.700	&	15.677	&	15.513	&	15.501	&	15.154	&	15.151	&	R	R	R	&	 \\ 
214	&	0.53952	&	RRab	&	15.994	&	15.911	&	15.741	&	15.706	&	15.323	&	15.305	&	R	R	R	&	1\\ 
215	&	0.528746	&	RRab	&	15.935:	&	15.885:	&	15.607	&	15.585	&	15.339:	&	15.324:	&	R	R	R	&	Bl\\ 
216	&	0.3464803	&	RRc	&	15.929	&	15.908	&	15.677	&	15.667	&	15.492	&	15.482	&	R	R	R	&	 \\ 
218	&	0.544862	&	RRab	&	16.108	&	16.075	&	15.708	&	15.658	&	15.249	&	15.244	&	R	R	R	&	Bl,c\\ 
219	&	0.6136073	&	RRab	&	16.011	&	15.993	&	15.717	&	15.707	&	 	&	 	&	R	R	 	&	 \\ 
220	&	0.6001119	&	RRab	&	15.977	&	15.946	&	15.724	&	15.710	&	15.205	&	15.197	&	R	R	R	&	Bl?,1\\ 
221	&	0.3787879	&	RRc	&	15.540	&	15.520	&	15.512	&	15.499	&	15.290	&	15.283	&	R	R	R	&	m,c\\ 
222	&	0.5967446	&	RRab	&	15.986	&	15.931	&	15.598	&	15.572	&	15.050	&	15.042	&	R	R	R	&	c,3\\ 
223	&	0.3291883	&	RRc	&	15.840	&	15.811	&	15.589	&	15.576	&	15.289	&	15.284	&	R	R	R	&	 \\ 
225	&	?	&	long	&	 	&	 	&	 	&	 	&	 	&	 	&	A	A	A	&	 \\ 
226	&	0.4884239	&	RRab	&	16.029	&	15.934	&	15.736	&	15.685	&	15.419:	&	15.378:	&	R	R	R	&	m\\ 
229	&	0.4991777	&	RRab	&	 	&	 	&	15.736	&	15.682	&	15.308:	&	15.289:	&	R	R	R	&	m,c\\ 
234	&	0.508039	&	RRab	&	16.046	&	15.963	&	15.685	&	15.645	&	15.445	&	15.428	&	R	R	R	&	c\\ 
235	&	0.759846	&	RRab	&	15.882	&	15.860	&	15.539	&	15.526	&	14.966	&	14.961	&	R	R	R	&	4\\ 
236	&	?	&	long	&	 	&	 	&	 	&	 	&	 	&	 	&	A	A	A	&	 \\ 
239	&	0.503982	&	RRab	&	15.907	&	15.815	&	15.754	&	15.700	&	15.282	&	15.258	&	R:	R:	R:	&	Bl,c,1\\ 
240	&	0.2760172	&	RRc	&	15.623	&	15.609	&	15.516	&	15.508	&	15.045	&	15.043	&	R:	R:	R:	&	 \\ 
241	&	0.596172	&	RRab	&	 	&	 	&	15.494	&	15.441	&	15.347	&	15.315	&	 	R	R	&	m\\ 
242	&	0.5964335	&	RRab	&	16.016	&	15.993	&	15.588	&	15.576	&	 	&	 	&	R	R	 	&	m.c\\ 
243	&	0.634627	&	RRab	&	15.911	&	15.891	&	15.640	&	15.629	&	15.012	&	15.007	&	R	R	R	&	Bl?,3\\ 
244	&	0.5378472	&	RRab	&	15.947:	&	15.840:	&	15.728	&	15.682	&	15.327:	&	15.309:	&	R	R	R	&	1\\ 
245	&	0.2840324	&	RRc	&	15.886	&	15.853	&	15.704	&	15.689	&	15.346	&	15.338	&	R	R	R	&	 \\ 
246	&	0.3391374	&	RRc	&	15.803	&	15.788	&	15.635	&	15.622	&	15.249	&	15.244	&	R	R	R	&	 \\ 
247	&	0.6053555	&	RRab	&	16.084	&	16.05	&	15.688	&	15.670	&	15.182	&	15.174	&	R	R	R	&	2\\ 
248	&	0.5097854	&	RRab	&	16.099	&	16.042	&	15.613	&	15.603	&	15.260	&	15.248	&	R	R	R	&	m\\ 
249	&	0.5330032	&	RRab	&	15.942	&	15.879	&	15.688	&	15.640	&	15.135	&	15.123	&	R	R	R	&	3\\ 
250n	&	0.5671499	&	RRab	&	 	&	 	&	15.523	&	15.506	&	 	&	 	&	 	R	 	&	m,c\\ 
250s	&	0.5861391	&	RRab	&	15.847	&	15.743	&	15.537	&	15.490	&	15.383	&	15.344	&	R	R	R	&	m,c\\ 
251	&	0.4705028	&	RRd	&	15.75:	&	15.714:	&	15.608	&	15.596	&	 	&	 	&	R	R	 	&	$P_1=0.348726$\\ 
252	&	0.3359896	&	RRd	&	 	&	 	&	15.728	&	15.719	&	 	&	 	&	R	R	 	&	$P_0=0.452791$\\ 
253	&	0.3326257	&	RRc	&	15.561:	&	15.543:	&	15.504	&	15.493	&	15.182	&	15.176	&	R	R	R	&	 \\ 
254	&	0.605656	&	RRab	&	15.794	&	15.772	&	15.545	&	15.531	&	15.200	&	15.193	&	R:	R	R	&	 \\ 
255	&	0.5726891	&	RRab	&	16.071	&	16.004	&	15.600	&	15.581	&	 	&	 	&	R:	R:	 	&	 \\ 
256	&	0.3180587	&	RRc	&	15.912	&	15.894	&	15.704	&	15.692	&	15.418	&	15.412	&	R	R	R	&	 \\ 
257	&	0.6019823	&	RRab	&	16.056	&	15.990	&	15.733	&	15.705	&	15.110:	&	15.089:	&	R	R	R	&	Bl?\\ 
258	&	0.713396	&	RRab	&	16.006	&	15.984	&	15.599	&	15.587	&	15.072	&	15.068	&	R	R	R	&	4\\ 
259	&	0.3335311	&	RRc	&	15.999	&	15.968	&	15.733	&	15.720	&	15.312	&	15.309	&	R	R	R	&	c\\ 
260	&	?	&	long	&	 	&	 	&	 	&	 	&	 	&	 	&	A	A	A	&	 \\ 
261	&	0.4448913	&	RRc(?)	&	15.422	&	15.413	&	15.195	&	15.187	&	15.110	&	15.107	&	R:	R:	R:	&	c\\ 
263	&	0.2168508	&	SXPhe	&	17.443	&	17.437	&	17.046	&	17.043	&	18.009	&	17.991	&	R	R	R	&	 \\ 
264	&	0.356466	&	RRc	&	15.622	&	15.599	&	15.474	&	15.465	&	15.143	&	15.139	&	R:	R:	R:	&	 \\ 
266	&	0.3424518	&	RRc	&	15.849	&	15.834	&	15.516	&	15.508	&	15.308	&	15.307	&	R:	R:	R:	&	c\\ 
269	&	0.3592861	&	RRc	&	15.600	&	15.574	&	15.510	&	15.505	&	15.090	&	15.093	&	R:	R:	R:	&	 \\ 
270n	&	0.625819	&	RRab	&	 	&	 	&	 	&	 	&	 	&	 	&	R	R	R	&	m,c\\ 
270s	&	0.690195	&	RRab	&	 	&	 	&	 	&	 	&	 	&	 	&	R	R	R	&	m,c\\ 
271	&	0.632794	&	RRab	&	16.031	&	15.997	&	15.640	&	15.624	&	15.182	&	15.175	&	R	R	R	&	3\\ 
272	&	?	&	long	&	 	&	 	&	 	&	 	&	 	&	 	&	A	A	A	&	 \\ 
273	&	46.43	&	long	&	 	&	 	&	 	&	 	&	 	&	 	&	A	A	A	&	 \\ 
274	&	?	&	long	&	 	&	 	&	 	&	 	&	 	&	 	&	A	A	A	&	 \\ 
N1	&	0.071628533	&	SXPhe	&	17.719	&	17.660	&	17.454	&	17.428	&	17.212	&	17.201	&	R	R	R	&	 \\ 
N2	&	0.251010	&	RRc	&	15.707	&	15.700	&	15.535	&	15.534	&	 	&	 	&	R:	R:	 	&	 \\ 
N3	&	0.29654065	&	RRc	&	15.688	&	15.681	&	15.547	&	15.544	&	15.426	&	15.425	&	R:	R:	R:	&	 \\ 

\\[-7pt]
\hline
\label{table_var_all}
\end{longtable}
\label{lastpage}


\begin{thebibliography}{}
\bibitem[\protect\citeauthoryear{Akerlof et al.}{2000}]{ROTSE}
Akerlof C., Amrose S., Balsano R. et al., 2000, AJ, 119, 1901
\bibitem[\protect\citeauthoryear{Alard}{2000}]{Al2}
Alard C., 2000, A\&AS, 144, 363
\bibitem[\protect\citeauthoryear{Alard \& Lupton}{1998}]{AlaLup}
Alard C., Lupton R. H., 1998, ApJ, 503, 325
\bibitem[\protect\citeauthoryear{Alcock et al.}{2000}]{Alcock2000}
Alcock C., Allsman R., Alves D. R., Axelrod T., Becker A., et al.,
2000, ApJ, 542, 257
\bibitem[\protect\citeauthoryear{Alcock et al.}{2003}]{Alcock}
Alcock C., Alves D. R., Becker A., Bennett D., Cook K. H. et al.,
2003, ApJ, 598, 597
\bibitem[\protect\citeauthoryear{Arellano Ferro et al.}{2002}]{AFer}
Arellano Ferro A., Aguilar A., Mar\'\i n Z., Rosenzweig P.,
2002, Rev. Mex. A\&A Conf. Ser., 14, 39
\bibitem[\protect\citeauthoryear{Bailey}{1913}]{Bailey}
Bailey S. I., 1913, Harvard Ann., 78, 1
\bibitem[\protect\citeauthoryear{Bakos}{2000}]{calib}
Bakos G. \'A., 2000, Occasional. Tech. Notes Konkoly Obs.,
No. 11, \url{http://www.konkoly.hu/Mitteilungen/Mitteilungen.html}
\bibitem[\protect\citeauthoryear {Bakos, Benk\H{o} \& Jurcsik}
{Bakos et al.}{2000}]{Acta}
Bakos G. \'A., Benk\H{o} J. M., Jurcsik J., 2000, Acta Astron., 50, 221
\bibitem[\protect\citeauthoryear{Benk\H{o}}{2001}]{Ben}
Benk\H{o} J. M., 2001, in Chen W-P, Lemme C. and Paczy\'nski B. (eds),
ASP Conf. Ser., 246, Small Telescope Astronomy in Global Scale, p. 329
\bibitem[\protect\citeauthoryear{Benk\H{o} \& Jurcsik}{2000}]{BJ}
Benk\H{o} J. M., Jurcsik J., 2000, in Szabados L. and Kurtz D. (eds),
ASP Conf. Ser., 203, The Impact of Large-Scale Surveys on Pulzating Star
Research, p. 257
\bibitem[\protect\citeauthoryear{Blazhko}{1907}]{Blazhko}
Blazhko S., 1907, AN, 173, 325
\bibitem[\protect\citeauthoryear{Buchler \& Koll\'ath}{2002}]{KB}
Buchler J. R., Koll\'ath Z., 2002, ApJ, 573, 324
\bibitem[\protect\citeauthoryear{Castellani, Castellani \& Cassisi}{Castellani et al.}{2005}]{Castellani}
Castellani M., Castellani V., Cassisi S., 2005, A\&A, 437, 1017
\bibitem[\protect\citeauthoryear{Castellani \& Tornamb\'e}{1981}]{CT}
Castellani V., Tornamb\'e A., 1981, A\&A, 96, 207
\bibitem[\protect\citeauthoryear{Catelan}{2004}]{Cat04}
Catelan M., 2004, ApJ, 600, 409
\bibitem[\protect\citeauthoryear{Catelan, Pritzl \& Smith}{Catelan et al.}{2004}]{Catelan}
Catelan M., Pritzl B. J., Smith H. A., 2004, ApJS, 154, 633
\bibitem[\protect\citeauthoryear{Cacciari, Corwin \& Carney}{Cacciari et al.}{2005}]{Cacciari}
Cacciari C., Corwin T. M., Carney B. W., 2005, AJ, 129, 267 (CC05)
\bibitem[\protect\citeauthoryear{Carretta et al.}{1998}]{C98}
Carretta E., Cacciari C., Ferraro F. R., Fusi Pecci F.,
Tessicini G., 1998, MNRAS, 298, 1005 (C98)
\bibitem[\protect\citeauthoryear{Clement \& Goranskij}{1997}]{V79}
Clement C. M., Goranskij V. P., 1997, ApJ, 513, 767
\bibitem[\protect\citeauthoryear{Clement \& Rowe}{2000}]{CR}
Clement C. M., Rowe J., 2000, AJ, 120, 1579
\bibitem[\protect\citeauthoryear{Clement, Ferance \& Simon}{Clement et al.}{1993}]{Clement93}
Clement C. M., Ferance S., Simon N. R., 1993, ApJ, 412, 183
\bibitem[\protect\citeauthoryear{Clement et al.}{1997}]{Clement97}
Clement C. M., Hilditch R. W., Kaluzny J., Rucinski S. M., 1997, ApJ, 489, L55
\bibitem[\protect\citeauthoryear{Clement et al.}{2001}]{Cl}
Clement C. M., Muzzin A., Dufton Q.,
Ponnampalam T., Wang J., Burford J., Richardson A., Rosebery T., Rowe J.,
Sawyer Hogg H., 2001, AJ, 122, 2587 (Catalogue)
\bibitem[\protect\citeauthoryear{Clementini et al.}{2004}]{CC04}
Clementini G., Corwin T. M., Carney B. W., Sumerel A. N., 2004, AJ, 127, 938 (CC04)
\bibitem[\protect\citeauthoryear{Corwin \& Carney}{2001}]{CC01}
Corwin T. M., Carney B. W., 2001, AJ, 122, 3183 (CC01)
\bibitem[\protect\citeauthoryear {Corwin, Carney \& Allen}
{Corwin et al.}{1999}]{CCRRd}
Corwin T. M., Carney B. W., Allen D. M., 1999, AJ, 117, 1332
\bibitem[\protect\citeauthoryear{Cudworth}{1979}]{Cud}
Cudworth M., 1979, AJ, 84, 1312
\bibitem[\protect\citeauthoryear{Dziembowski \& Mizerski}{2004}]{Dziem}
Dziembowski W. A., Mizerski T., 2004, Acta Astron., 54, 363 
\bibitem[\protect\citeauthoryear{Greenstein}{1935}]{Greenstein} 
Greenstein J. L., 1935, AN, 257, 301
\bibitem[\protect\citeauthoryear{Guhathakurta et al.}{1994}]{Gua}
Guhathakurta P., Yanny B., Bahcall J. N., Schneider D. P., 1994, AJ, 108, 1786
\bibitem[\protect\citeauthoryear{Hartman et al.}{2005}]{hartman}
Hartman J. D., Kaluzny J., Szentgyorgyi A., Stanek K. Z., 2005, AJ, 129, 1596
\bibitem[\protect\citeauthoryear{Jerzykiewicz, Schult \& Wenzel}{Jerzykiewicz et al.}{1982}]{Jerzy}
Jerzykiewicz M., Schult, R. H., Wenzel, W., 1982, Acta Astron., 32, 357
\bibitem[\protect\citeauthoryear{Jurcsik \& Barlai}{1990}]{JurcsikBarlai}
Jurcsik J., Barlai K., 1990, in Cacciari C. and  Clementini G. (eds), ASP Conf. Ser. 11., 
Confrontation between Stellar Pulsation and Evolution, p. 112
\bibitem[\protect\citeauthoryear{Jurcsik et al.}{2003}]{letter}
Jurcsik J., Benk\H{o} J. M., Bakos G. \'A., Szeidl B., Szab\'o R., 2003, ApJ, 597, L49
\bibitem[\protect\citeauthoryear{Jurcsik et al.}{2005a}]{RRGem}
Jurcsik J., S\'odor \'A., V\'aradi M., Szeidl B., Wash\"uttl M., et al., 
2005a, A\&A, 430, 1049 
\bibitem[\protect\citeauthoryear{Jurcsik et al.}{2005b}]{HannaActa}
Jurcsik J., Szeidl B., Nagy A., S\'odor \'A., 2005b, Acta Astron., 55, 303
\bibitem[\protect\citeauthoryear{Kaluzny et al.}{1998}]{K98} 
Kaluzny J., Hilditch R. W., Clement C., Rucinski S. M.,
1998, MNRAS, 296, 347 (K98)
\bibitem[\protect\citeauthoryear{Koll\'ath}{1990}]{Zoli}
Koll\'ath Z., 1990, Occasional Tech. Notes Konkoly Obs. No. 1
 \url{http://www.konkoly.hu/staff/kollath/mufran.html}
\bibitem[\protect\citeauthoryear{Kukarkin \& Kukarkina}{1961}]{Kuk61} 
Kukarkin B. V., Kukarkina, N. P., 1961, Perem. Zvezdy, 13, 309 
\bibitem[\protect\citeauthoryear{Kukarkin \& Kukarkina}{1970}]{Kuk70} 
Kukarkin B. V., Kukarkina, N. P., 1970, Perem. Zvezdy, 17, 157 
\bibitem[\protect\citeauthoryear{Larink}{1922}]{Larink}
Larink J., 1922, Astron. Abhang. Hamburger Sternw. Bergedorf, 2, No.~6 
\bibitem[\protect\citeauthoryear{Layden \& Sarajedini}{2003}]{Layden}
Layden, A. C., Sarajedini, A., 2003, AJ, 125, 208 
\bibitem[\protect\citeauthoryear{Mallik, Christensen \& Saha}{Mallik et al.}{1999}]{Mallik} 
Mallik P., Christensen J., Saha A., 1999, BAAS, 31, 1481
\bibitem[\protect\citeauthoryear{Marconi et al.}{2003}]{Marconi} 
Marconi M., Caputo F., Di Criscienzo M., Castellani M., 2003, ApJ, 596, 299
\bibitem[\protect\citeauthoryear{Moskalik \& Poretti}{2003}]{Moskalik} 
Moskalik  P., Poretti E., 2003, A\&A, 398, 213
\bibitem[\protect\citeauthoryear{Murtagh \& Heck}{1987}]{Murtagh} 
Murtagh F., Heck A., 1987, Multivariate Data Analysis, Riedel, Dordrecht
\bibitem[\protect\citeauthoryear{M\"uller}{1933}]{Mueller} 
M\"uller Th., 1933, Ver\"off. Univ. Berlin-Babelsberg, 11, No.~1
\bibitem[\protect\citeauthoryear{Nemec \& Clement}{1989}]{NC89} 
Nemec J. M.,  Clement C. M., 1989, AJ, 98, 860 
\bibitem[\protect\citeauthoryear{Oaster, Smith \& Kinemuchi}{Oaster \& al.}{2006}]{Oaster} 
Oaster L., Smith H. A., Kinemuchi K., 2006, PASP, 118, 405
\bibitem[\protect\citeauthoryear{Olech et al.}{1999a}]{Olech99a} 
Olech A., Kaluzny J., Thompson I. B., Pych W., Krzeminski W., 
Schwarzenberg-Czerny A., 1999a, AJ, 118, 442  
\bibitem[\protect\citeauthoryear{Olech et al.}{1999b}]{Olech99b} 
Olech A., Woz\'niak P. R., Alard C., Kaluzny J., Thompson I. B.,
1999b, MNRAS, 310, 759
\bibitem[\protect\citeauthoryear{Olech et al.}{2001}]{Olech01} 
Olech A., Kaluzny J., Thompson I. B., Pych W.,  Krzeminski W., 
Schwarzenberg-Czerny A., 2001, MNRAS, 321, 421
\bibitem[\protect\citeauthoryear{Oosterhoff}{1939}]{Oo}
Oosterhoff P. Th., 1939, Observatory, 62, 104
\bibitem[\protect\citeauthoryear{Panov}{1980}]{Panov} 
Panov K., 1980, Perem. Zvezdy, 21, 391
\bibitem[\protect\citeauthoryear{Purdue et al.}{1995}]{Purdue}
Purdue P., Silbermann N. A., Gay P., Smith H. A., 1995, AJ, 110, 1712
\bibitem[\protect\citeauthoryear{Press et al.}{1992}]{NumRec} 
Press W. H., Teukolsky S. A., Vetterling W. T.,  
Flannery B. P., 1992, Numerical Recipes, 2nd edn. 
Cambridge Univ. Press, Cambridge
\bibitem[\protect\citeauthoryear{Pritzl et al.}{2003}]{Pritzl} 
Pritzl B. J., Smith H. A., Stetson P. B., 
Catelan M., Sweigart A. V., Layden A. C., Rich M. R., 2003, AJ, 126, 1381
\bibitem[\protect\citeauthoryear{Roberts \& Sandage}{1955}]{RS} 
Roberts M., Sandage A. R., 1955, AJ, 60, 185
\bibitem[\protect\citeauthoryear{Rood \& Crocker}{1989}]{RC}
Rood R. T., Crocker D. A., 1989, in Schmidt E. G. (ed.), IAU Colloq. 111,  
The Use of Pulsating Stars in Fundamental Problems of Astronomy, 
 Cambridge Univ. Press, p. 103 
\bibitem[\protect\citeauthoryear{Schechter, Mateo \& Saha}{Schechter et al.}{1993}]{DoP} 
Schechter P. L., Mateo M., Saha A., 1993, PASP, 105, 1342
\bibitem[\protect\citeauthoryear{Smith}{1995}]{Smith} 
Smith H. A., 1995, RR Lyrae Stars, Cambridge Univ. Press
\bibitem[\protect\citeauthoryear{Soszy\'nski et al.}{2003}]{Soszy} 
Soszy\'nski I.,  Udalski A., Szyma\'nski M., Kubiak M., Pietrzy\'nski G., 
Wo\'zniak P., \.Zebru\'n K., Szewczyk O., Wyrzykowski L.,  
2003, Acta Astron., 53, 93
\bibitem[\protect\citeauthoryear{Stellingwerf}{1978}]{Stel} 
Stellingwerf R. F., 1978, ApJ, 224, 953
\bibitem[\protect\citeauthoryear{Stetson}{1987}]{daophot} 
Stetson P. B., 1987, PASP, 99, 191 
\bibitem[\protect\citeauthoryear{Stetson}{2000}]{St3}
Stetson P. B., 2000, PASP, 112, 925
\bibitem[\protect\citeauthoryear{Strader, Everitt \& Danford}
{Strader et al.}{2002}]{Strader}
Strader J., Everitt, H. O., Danford S., 2002 MNRAS, 335, 621 (S02)
\bibitem[\protect\citeauthoryear{Szab\'o, Koll\'ath \& Buchler}{Szab\'o et al.}{2004}]{Robi} 
Szab\'o R., Koll\'ath Z., Buchler J. R., 2004 A\&A, 425, 627 
\bibitem[\protect\citeauthoryear{Szeidl}{1965}]{Bela65} 
Szeidl B., 1965, Comm. Konkoly Obs. Budapest, No 58
\bibitem[\protect\citeauthoryear{Szeidl}{1973}]{Bela73} 
Szeidl B., 1973, Comm. Konkoly Obs. Budapest, No 63
\bibitem[\protect\citeauthoryear{Tucholke, Scholz \& Brosche}
{Tucholke et al.}{1994}]{Tu} 
Tucholke J., Scholz R.-D., Brosche P., 1994, A\&AS, 104, 161
\bibitem[\protect\citeauthoryear{Uglesich et al.}{2000}]{Ug} 
Uglesich R., Mirabal N., Sugerman B. and Crotts A., 2000, BAAS, 32, 879
\bibitem[\protect\citeauthoryear{van Albada \& Baker}{1973}]{vAB} 
van Albada T. S., Baker N., 1973, ApJ, 185, 477
\bibitem[\protect\citeauthoryear{Walker}{1994}]{Walker} 
Walker A., 1994, AJ, 108, 555
\bibitem[\protect\citeauthoryear{Welty}{1985}]{Welty} 
Welty D. E., 1985, AJ, 90, 2555
\end{thebibliography}
\end{document}